%% file: kawabata-MS4.tex
\documentclass[preprint2]{aastex}
\usepackage{hyperref}
\usepackage{cleveref}
\usepackage{afterpage,bm,spr-astr-addons}
\RequirePackage{color}
\shorttitle{Astropys. Space Sci. }
\shortauthors{Kawabata}

\begin{document}

\title{15-Digit Accuracy Calculations of Ambartsumian-Chandrasekhar's $\bm{H}$-Functions
 for Four-Term Phase Functions with the Double-Exponential Formula\ \ \ \ \ \ \ {\small \bf (To be published in Astrophysics and Space Science, January 2018, 363:1)}  }


\author{Kiyoshi Kawabata\altaffilmark{}}
\affil{Department of Physics,  Tokyo University of Science,
    Shinjuku-ku, Tokyo 162-8601, Japan\\
    E-mail: kawabata@rs.kagu.tus.ac.jp}

\email{kawabata@rs.kagu.tus.ac.jp}

\begin{abstract}
We have established an iterative scheme to calculate with 15-digit accuracy the numerical values of Ambartsumian-Chandrasekhar's $H$-functions for anisotropic scattering characterized by the four-term phase function: the method incorporates 
  some  advantageous features of the iterative procedure of Kawabata (2015, Astrophys. Space Sci.~358:32)  and the double-exponential integration formula~(DE-formula) of Takahashi and Mori (1974, Publ. RIMS, Kyoto Univ. 9, 721), which proved highly effective  in  Kawabata (2016, Astrophys. Space Sci.~361:373).
Actual calculations of the $H$-functions have been carried  out employing   27 selected cases of  the  phase function, 56 values of the single scattering albedo $\varpi_0$, and 36 values of an angular variable $\mu(=\cos \theta)$, with $\theta$ being the zenith angle specifying the direction of  incidence  and/or emergence of radiation.
 Partial results obtained  for  conservative isotropic
scattering, Rayleigh scattering, and  anisotropic scattering due to  a full four-term  phase function   are presented.
They indicate that it is important to simultaneously verify accuracy  of the numerical values of the $H$-functions for $\mu <0.05$, the domain often neglected in tabulation.
As a sample application of the isotropic scattering $H$-function,
an attempt is made in Appendix to  simulate by iteratively solving the Ambartsumian equation the values of the plane and spherical albedos of a semi-infinite, homogeneous atmosphere calculated  by  Rogovtsov and Borovik (2016, J. Quant. Spectr. Radiat. Transf. 183, 128), who employed  their analytical representations for these quantities and   the single-term  and two-term Henyey-Greenstein phase functions of appreciably   high degrees of anisotropy. 
While our results are in satisfactory  agreement with theirs,  our procedure is in need of  a faster  algorithm   to  routinely deal with  
problems involving  highly  anisotropic phase functions giving rise to near-conservative scattering.
\end{abstract}


%%%%%%%%%%%%%%%%%%%%%%%%%%%%%%%%%%%%%%%%%%%%%%%%%%%%%%%%%%%%%
\section{Introduction}
%%%%%%%%%%%%%%%%%%%%%%%%%%%%%%%%%%%%%%%%%%%%%%%%%%%%%%%%%%%%%
Let $I_r(\mu, \phi)$ be the intensity of radiation diffusely reflected into the direction $(\mu, \phi)$ by a semi-infinite plane-parallel atmosphere
illuminated by mono-directional sunlight coming from a direction $(\mu_0, \phi_0)$ with the flux $\pi F_0$ per unit area perpendicular to the incident beam: here, $\mu$ and $\phi$ are respectively the cosine of zenith angle $\theta$ and azimuth angle, while $\mu_0(=\cos \theta_0)$  and $\phi_0$ are the similar quantities specifying the incident direction \citep[see, e.g.,][]{han74, kaw80}.
Then $I_r(\mu, \phi)$ can be expressed in terms of a reflection function $R(\mu, \mu_0, \phi-\phi_0)$ as 
\begin{multline}
I_r(\mu, \phi)=\mu_0R(\mu, \mu_0, \phi-\phi_0)F_0,   \\ \label{eq-1}
   (0\le \mu, \mu_0 \le 1, -\pi\le \phi-\phi_0 \le\pi),
\end{multline}
provided that we can ignore the effect of polarization of light as we shall assume throughout this work.
The azimuth angle dependence of the reflection function  is usually taken care of by applying the Fourier series expansion: 
\begin{multline}
R(\mu, \mu_0, \phi-\phi_0) \\
=\sum_{m=0}^\infty (2-\delta_{m0})R^{(m)}(\mu, \mu_0)\cos m(\phi-\phi_0), \label{eq-2}
\end{multline}
where
the symbol $\delta_{0k}$ signifies  the Kronecker delta
such that 
\begin{equation}
\delta_{0k}=\begin{cases}
                 1  &  \text{if $k=0$}, \\
                 0  &  \text{otherwise}.
                \end{cases}  \label{eq-3}
\end{equation}
The calculation of $I_r(\mu, \phi)$ therefore boils down to finding the values of the Fourier coefficients of reflection function $R^{(m)}(\mu, \mu_0)$. One of the  most straightforward procedures would be to solve  Ambartsumian's equation as shown  by Eq.(A1) in  Appendix of the present work  for $R^{(m)}(\mu, \mu_0)$ (see, e.g., Ambartsumian 1958; Hansen and Travis 1974; Sobolev 1975; Goody and Yung 1989; Yanovitskij 1997; Mishchenko \textit{et al.} 2006).
%\citep[see, e.g.,][]{amb58, han74, sob75, god89, yan97, mis06}. 
A sophisticated iterative procedure to numerically solve this equation was developed by \cite{mis99}. \par
Furthermore,  a very powerful  method of solving wide ranges of problems of radiative transfer has been constructed  by \cite{rog16}\citep[see also][and references therein]{rog09, rog10} based on application of  general invariance relations, where new analytical representations for reflection functions and other related quantities are given. Judging from their results obtained for semi-infinite plane-parallel atmospheres, the method appears to be especially advantageous in dealing with  highly anisotropic phase functions.
 \par  
 The Fourier coefficient  $R^{(m)}(\mu, \mu_0)$
  is often expressed in the following form particularly for numerical calculations:
\begin{align}
R^{(m)}(\mu, \mu_0)&=\frac{\varpi_0}{4(\mu+\mu_0)}H^{(m)}(\varpi_0, \mu)H^{(m)}(\varpi_0, \mu_0)\times \notag \\
&\times V^{(m)}(\mu, \mu_0), \label{eq-4}
\end{align}
where  $\varpi_0$ is the single scattering albedo, $H^{(m)}(\varpi_0, \mu)$ is the
 Ambartsumian-Chandrasekhar $H$ function\citep{vii93} or simply the Chandrasekhar $H$ function (Kolesov and Smoktii 1972),  and $V^{(m)}(\mu, \mu_0)$ is a polynomial function of two variables \citep[][]{sob75, hul80}.
This  function $H^{(m)}(\varpi_0, \mu)$ is known to satisfy the following Ambartsumian-Chandrasekhar equation:    
\begin{eqnarray}
H^{(m)}(\varpi_0,\mu)&=&1+\mu  H^{(m)}(\varpi_0,\mu) \nonumber \\
& &\hspace{-1cm} \times\int_0^1\!\!\frac{\psi^{(m)}(\mu')}{\mu+\mu'}H^{(m)}(\varpi_0,\mu')d\mu'.  \label{eq-5} 
\end{eqnarray}
where the function $\psi^{(m)}(\mu)$  is the $m$-th order Fourier component of the characteristic function derived from the  phase function of our interest. It must be noteworthy that Eq.\eqref{eq-5}  was iteratively solved for the first time by \cite{amb43a} for the case of isotropic scattering.
 A general recipe to calculate $\psi^{(m)}(\mu)$ is shown by \cite{sob75} and \cite{hul80}. 
As mentioned in \cite{iva73}, Eq.\eqref{eq-5} was  derived originally by \cite{hal38} in consideration of multiple, isotropic scattering of neutrons, and later independently by means of use of formal mathematical transformations or properties of invariance by \cite{amb42, amb43a, amb43b, amb58}
\citep[see also][]{sob63, sob75} to solve the problems of diffuse reflection of light by isotropically scattering, homogeneous semi-infinite media. Later, \cite{cha50} arrived at  Eq.\eqref{eq-5} based on  consideration of the similar problems for mildly anisotropic scattering phase functions. This equation was  then derived in more rigorous manner by \cite{sob63, sob75}.  For historical background and discussion on the mathematical properties of of Eq.\eqref{eq-5}, we refer the reader to, e.g.,   
\cite{cha50}, \cite{sti59}; \cite{kou63}, \cite{iva73}, \cite{sob75}, \cite{hul80}, \cite{vii93}, and \cite{yan97}. \par
The $H$-functions and their derivatives with respect to angular variable and  single scattering albedo  are important  not only for  the theory of radiative transfer\citep{hul88, hov88, vii93} but also in other disciplines of physical sciences such as the electron transports in condensed matter physics \citep[see, e.g.,][and the references cited therein]{mon08,jab12,jab15}, 
so that   numerical values  of high accuracy (better than 10 significant digits) of the $H$-functions, in  particular for  isotropic scattering, are required\citep[][]{jab15}.  In fact, a great deal of efforts have been devoted by various investigators
to exploiting  accurate and yet efficient methods to numerically evaluate the $H$-functions.\par
\cite{sob75} develops a theory that  potentially allows one to obtain  formal analytical representations for the $H$-functions for arbitrary phase functions.
For  isotropic scattering, a variety of  real integral representations for the $H$-function  are available including the one derived by \cite{hop34}
\citep[see, e.g., ][]{bus54, sti59,  kou63, iva73, hul80, rut87}.\par
  Nevertheless, it is not  straightforward to accurately evaluate the $H^{(m)}(\varpi_0, \mu)$ and requires some ingenuity to attain the  15-digit accuracy permitted by the double precision arithmetic even in the simplest case of isotropic scattering.
Of particular interest from this view point is the work done by \cite{vii86}, who successfully obtained the values of the $H$-functions with 14-digit accuracy or better for the cases of isotropic, Rayleigh, and linearly anisotropic scattering laws by approximating Sobolev's resolvent function using exponent series.\par
Inspired by this, \cite{kaw11} calculated the values of the $H$-function for  isotropic scattering  on a fine mesh of  $(\varpi_0, \mu)$, employing the real integral representation  given by \cite{rut87}: they circumvented the numerical difficulty in integration  that arises near  the origin  of the integration variable $x$  by the sum of an approximate analytical integration over a small interval $x\in[0, \varepsilon]$  and  a numerical integration over the remaining interval $x\in[\varepsilon, \pi/2]$ obtained by  the Gauss-Legendre quadrature, to achieve 11-digit accuracy.  Simultaneously, they carried out  a least-squares fit to their results, to produce a rational approximation formula for $H(\varpi_0, \mu)$ for isotropic scattering, whose maximum relative error is 
 supposed to be  $2.1\times 10^{-6}(=2.1\times 10^{-4} \%)$. 
However, it was found  hard to further upgrade with their  scheme the numerical accuracy  of evaluating  the $H$-function.\par
 \cite{bos83}, on the other hand,  had succeeded in getting  the values of  $H^{(m)}(\varpi_0, \mu)$
  for isotropic, linearly anisotropic, and Rayleigh scatterings with 11-digit accuracy by iteratively solving an alternative form of Eq.\eqref{eq-5}   derived by \cite{cha50}\citep[see also][]{sob75}: 
%%%%%%%%%%%%%%%%%%%%%%%%%%%%%%%%%%%%%%%%%%%%%%%%%%%%%%
\begin{equation}
H^{(m)}(\varpi_0,\mu)=\left\{\sqrt{1-2\psi^{(m)}_0}
+S^{(m)}(\mu)\right\}^{-1},  \label{eq-6}
\end{equation} 
%%%%%%%%%%%%%%%%%%%%%%%%%%%%%%%%%%%%%%%%%%%%%%%%%%%%%%
with  
\begin{subequations}
\begin{align}
\psi^{(m)}_0&=\int_0^1\!\!\!\psi^{(m)}(\mu')d\mu',   \label{eq-7a} \\
S^{(m)}(\mu)&=\displaystyle{\int_0^1\!\!F^{(m)}(\mu,\mu^\prime)d\mu^\prime,   } \label{eq-7b} \\
F^{(m)}(\mu, \mu^\prime)&=\displaystyle{\frac{\mu'\psi^{(m)}(\mu')H^{(m)}(\varpi_0,\mu')}{\mu+\mu'}}. \label{eq-7c}
\end{align} 
\label{eq-7}
\end{subequations} 
\hspace*{-0.2cm}They employ the 128-point Gauss-Legendre quadrature  to carry out the required integrations with respect to $\mu^\prime$, and the iteration is   initiated by setting $H^{(m)}(\varpi_0, \mu)=1$. The resulting set of new values for $H^{(m)}(\varpi_0, \mu)$ is  normalized by the value of $H^{(m)}(\varpi_0, 0)$ before entering the next iterative process. To refine the iterate, a weighted mean of  the  newly obtained  and  the one that precedes is taken  as a new starting value  for the next iterative step.  In so doing, they achieved 11-digit accuracy
 for all the cases.   \par
\cite{kaw15} performed calculations of  the 
values of $H^{(m)}(\varpi_0, \mu)$ for some of the four-term phase functions basically
following  the method of  Bosma and de Rooij (1983) except that the starting approximation for the iteration was constructed by the approximate formula of  \cite{kaw11} for the azimuth angle independent ($m=0$) components, and values of $H^{(m-1)}(\varpi_0, \mu)$ for $m\ge 1$.   \par
\cite{nat97} (see also Mohankumar and Natarajan 2008)
%\citep[see also ][]{mon08} 
developed another  numerical scheme to evaluate the values of $H(\varpi_0,$ $\mu)$ for isotropic scattering by way of  the related $X$ functions of neutron transport employing two modified trapezoidal quadrature schemes. Their results are supposed to be accurate to 14 digits, although unfortunately no numerical results are given.  \par
\cite{jab15} has proposed a scheme that incorporates various integral representations for the solution of the isotropic scattering $H$-function, enabling him to obtain numerical values with accuracy of 21 digits or higher
in quadruple  precision arithmetic (private communication).  \par
By means of the DE-formula of \cite{tm74}~\citep[see also][]{ms01},  whose optimality is mathematically proven by \cite{sug97},  \cite{kaw16} repeated his foregoing   calculations \citep{kaw11} of the isotropic scattering $H$-function,
and demonstrated that it is possible to get  numerical values with accuracy of 15 digits in double precision arithmetic.  In fact, the results are in perfect agreement within one unit difference in the 15-th decimal place with those of \cite{jab15}.  \par
The principal purpose of the present work is  to upgrade the iterative scheme  of  \cite{kaw15} by incorporating an  automatic error-control capability into  the  DE-formula, to evaluate the $H$-functions 
with the accuracy permitted by the double-precision calculations. 
We shall restrict our consideration to the four-term phase function as in \cite{kaw15}
in view of the remarks given by  \cite{hul80}
that  the $H$-function method as a means to obtain reflection functions is not of practical use for phase functions more complex than that. \par
For realistic phase functions,  a more efficient way to get the reflection functions  for semi-infinite, homogeneous media would conceivably be to  solve  the Ambartsumian  equation as has been carried out  by  \cite{mis99} or  to rely  on  totally  different procedures such as the analytical representations for reflection function and the corresponding plane and spherical albedos derived  by \cite{rog16} using  the special Fredholm linear integral equations for reflection function and its azimuthal harmonics. \par
In view of the above and as an application of the isotropic scattering $H$-function, we shall solve in our Appendix the Ambartsumian  equation by  a straightforward successive approximation  in  an attempt to reproduce the values of the plane and spherical albedos calculated by \cite{rog16}, who used  their analytical expressions  for these quantities assuming  the single-term  and two-term Henyey-Greenstein phase functions with six values for the anisotropy parameter $g$, viz., $0.989$, $0.99$, $0.989$, $\pm 0.995$, and $0.9965$,  and 16 values ranging from $0.5$ to $0.9999$ for the single scattering albedo $\varpi_0$.%%%%%%%%%%%%%%%%%%%%%%%%%%%%%%%%%%%%%%%%%%%%%%%%%%%%%%%%
\section{Formalism} 
%%%%%%%%%%%%%%%%%%%%%%%%%%%%%%%%%%%%%%%%%%%%%%%%%%%%%%%%
\subsection{Characteristic Function Employed}
For the purpose of testing the range of applicability of the scheme developed in the present work, 
we shall employ, as in \cite{kaw15}, a four-term phase function of the form
\begin{equation}
P(\Theta)=\varpi_0\sum_{m=0}^{M}x_mP_m(\cos\Theta)  \quad (M\le 3),\label{eq-8}
\end{equation}
where $P_m(\cos\Theta)$ is the Legendre polynomial function of the $m$-th degree, $x_m$'s  are the expansion coefficients with $x_0$ being fixed to  unity, and $\Theta$ is the scattering angle, 
while $M$ is the highest degree of the Legendre functions to be taken into account, and coincides with the highest degree of the Fourier terms required to represent the azimuth-angle dependence of reflection function (see Eq.\eqref{eq-2}), which is in turn the  highest degree of the corresponding Fourier components  $H^{(m)}(\varpi_0,\mu)$ and $\psi^{(m)}(\mu)$ to be taken for Eq.\eqref{eq-4}. \par
 The four-term phase function, for which the $H$-function method is still feasible as a computational tool to obtain reflection functions for semi-infinite homogeneous atmospheres\citep{hul80},
  is convenient for  the fact that it covers,  as special cases, 
\begin{enumerate}
\item[(i)] isotropic scattering phase function  \\ 
with $x_m=0 \quad (m=1, 2,  3)$, 
\item[(ii)] Rayleigh scattering phase function with $x_1=0$, $x_2=1/2$, and $x_3=0$, 
\item[(iii)] linearly anisotropic scattering phase functions with  $x_1\ne 0$, and $x_2=x_3=0$,
\item[(iv)] three-term phase functions with $x_2\ne 0$, and $x_3=0$, 
\item[(v)] four-term phase functions with $x_3\ne 0$.
\end{enumerate} 
The Fourier components $\psi^{(m)}(\mu)\quad (m=0, 1, 2, 3)$ of the characteristic function derived from  the four-term  phase function Eq.\eqref{eq-8}  are  as follows: 
\begin{subequations}
\begin{align}
m=0&:\quad\psi^{(0)}(\mu)=\frac{1}{2}\varpi_0\left\{1+\frac{1}{4}x_2+\left(h_0x_1-\frac{3}{4}x_2  \right.\right. \notag \\ 
  &\left.-\frac{1}{4}h_0h_1x_2 +h_0x_3+\frac{1}{4}h_2x_3\right)\mu^2  \notag  \\
  &+\left(\frac{3}{4}h_0h_1x_2-\frac{5}{3}h_0x_3-\frac{5}{12}h_2x _3 \right. \notag \\
  &\left.\left. -\frac{1}{4}h_0h_1h_2x_3\right)\mu^4+\frac{5}{12}h_0h_1h_2x_3\mu^6\right\},     \label{eq-9a} \\
m=1&:\quad\psi^{(1)}(\mu)=\frac{1}{2}\varpi_0(1-\mu^2)\left\{\frac{1}{2}x_1+\frac{3}{16}x_3 \right. \notag   \displaybreak\\
\phantom{m=1:}  &+\left(\frac{1}{2}h_1x_2-\frac{1}{16}(h_1h_2+15)x_3\right)\mu^2\notag \\
  &\left.+\frac{5}{16}h_1h_2x_3\mu^4\right\},    \\
m=2&:\quad\psi^{(2)}(\mu)=\frac{3}{16}\varpi_0(1-\mu^2)^2 \notag\\
  &\times(x_2+x_3h_2\mu^2),   \\
m=3&:\quad\psi^{(3)}(\mu)=\frac{5}{32}\varpi_0x_3(1-\mu^2)^3, 
\end{align} \label{eq-9}
\end{subequations}   
\hspace*{-0.3cm}where $h_k$ is given by 
\begin{equation}
h_k=2k+1-\varpi_0x_k, \ \ (k=0, 1, 2,3) \label{eq-10}
\end{equation} 
as shown, e.g., by \cite{sob75} and \cite{hul80}
\footnote{It should be noted that the last term of Eq.(6a)  of \cite{kaw15}, which corresponds to Eq.\eqref{eq-9a} of the present work,  misses a multiplicative factor $x_3$ on account of a typographical error.
Furthermore,  the quantity $h_k$ was erroneously  referred to as the $k$-th moment of $H^{(m)}(\varpi_0, \mu)$ as indicated by his Eq.(7) by oversight.  However, all the discussions and the  numerical results presented therein remain valid  due
to the fact that all the calculations were performed using  the correct version of  equations, viz., Eqs.\eqref{eq-9a} and \eqref{eq-10}.}.\par
Substituting the characteristic functions $\psi^{(m)}(\mu^\prime)$ \quad $ (m=0, 1, 2, 3)$ given by Eqs.\eqref{eq-9} into Eq.\eqref{eq-7a} and carrying out integrations  analytically with respect to $\mu^\prime$,   we get the following results  for the expression inside the radical  sign  on the right-hand side of Eq.\eqref{eq-6}:  
\begin{subequations}
\begin{align}
m=0:&\quad  1-2\psi_0^{(0)}=\delta-\varpi_0\left\{\frac{1}{4}x_2+\frac{1}{3}\left(h_0x_1  \phantom{\frac{1}{3}}\right. \right.\notag\\ 
&\left.-\frac{3}{4}x_2-\frac{1}{4}h_0h_1x_2 +h_0x_3+\frac{1}{4}h_2x_3\right) \notag  \\
&+\frac{1}{5}\left(\frac{3}{4}h_0h_1x_2-\frac{5}{3}h_0x_3-\frac{5}{12}h_2x_3 \right. \notag\\
&\left.\left. -\frac{1}{4}h_0h_1h_2x_3\right)+\frac{5}{84}h_0h_1h_2x_3\right\},   \label{eq-11a}  \linebreak\\
m=1:&\quad 1-2\psi_0^{(1)}=1-\varpi_0\left\{\frac{1}{3}x_1+\frac{1}{8}x_3 \right. \notag \\
&+\frac{1}{15}\left[h_1x_2-\frac{1}{8}(h_1h_2+15)x_3\right] \notag \\
&\left.+\frac{1}{56}h_1h_2x_3\right\},   \label{eq-11b}  \\
m=2&:\quad1-2\psi_0^{(2)}=1-\frac{1}{5}\varpi_0\left(x_2+\frac{1}{7}x_3h_2\right), \hfill \\
m=3&:\quad 1-2\psi_0^{(3)}=1-\frac{1}{7}\varpi_0 x_3,
\end{align} \label{eq-11}
\end{subequations}
\hspace*{-0.1cm}where we have written $\delta=1-\varpi_0$, which proves useful to minimize the loss of significant digits when the value of $\varpi_0$ is close to unity\citep[see Eq.(5.59) of ][for more concise  form for Eqs.\eqref{eq-11}]{sob75}.\par
%%%%%%%%%%%%%%%%%%%%%%%%%%%%%%%%%%%%%%%%%%%%%%%%%%%%%%%%%%%%%%% 
\subsection{Numerical Integration Using  DE-Formula}
%% %%%%%%%%%%%%%%%%%%%%%%%%%%%%%%%%%%%%%%%%%%%%%%%%%%%%%%%%%%%%%
Application of the DE-formula to the integral involved in Eq.\eqref{eq-7b} requires the following variable transformations:
\begin{subequations} 
\begin{align}
\mu^{\prime\pm}\ \ &=\displaystyle{\frac{1\pm \phi(\xi)}{2}\quad \quad \quad\quad\quad\quad \ \  (0 \le \phi(\xi)\le 1),}  \label{eq-12a}\\
\phi(\xi)&=\displaystyle{\tanh\left[(\pi/2)\sinh(\xi)\right]~\quad\quad \ \  (0\le\xi<\infty),} \label{eq-12b}
\end{align}
\end{subequations}
where $\mu^{\prime+}$ corresponds to the case with $+\phi(\xi)$, and $\mu^{\prime-}$ to
 $-\phi(\xi)$ in Eq.\eqref{eq-12a}.
For a given value of $\mu$, we  then have 
\begin{multline}
\displaystyle{S^{(m)}(\mu)=\frac{\pi}{4}\int_0^\infty \!\!\! 
\left(\frac{\cosh(\xi)}{\cosh^2\left[(\pi/2)\sinh(\xi)\right]}\right) } \\
  \times Z\left[\mu, \phi(\xi)\right]d\xi, \label{eq-13} 
\end{multline}
where $Z[\mu, \phi(\xi)]$ is defined as
\begin{multline}
Z[\mu, \phi(\xi)]=F^{(m)}\left(\mu, \mu^{\prime+}\right)+F^{(m)}\left(\mu, \mu^{\prime-}\right)\\
=\displaystyle{F^{(m)}\left(\mu, \frac{1+\phi(\xi)}{2}\right)+F^{(m)}\left(\mu, \frac{1-\phi(\xi)}{2}\right).}    \label{eq-14}
\end{multline}
For  simplicity, let us  carry on our discussion for the time being under the assumption  that the function $Z[\mu, \phi(\xi)]$ can be evaluated  at will for arbitrary values of $\xi$.
Applying the trapezoidal rule, with division points $\xi_k=hk\quad(k=0, 1, \cdots)$ given with  a constant step-size $h$, to the integral
 on the right-hand side of Eq.\eqref{eq-13},  we get 
\begin{equation}
S^{(m)}(\mu)\simeq \displaystyle{
\frac{\pi h}{4}\left\{\sum_{k=\infty}^0\frac{2-\delta_{0k}}{2}w(kh)Z[\mu, \phi(kh)]\right\}, }  \label{eq-15}
\end{equation}
where
$w(kh)$ is the quadrature weight defined by 
\begin{equation}
w(kh)= \displaystyle{
\frac{\cosh(kh)}{\cosh^2\left[(\pi/2)\sinh(kh)\right]}, }  \label{eq-16}
\end{equation}  
whose calculation can be carried out efficiently by making use of the recurrence relation derived by \cite{wat90}.
In actual calculations, however, we need to  truncate the series in Eq.\eqref{eq-15} 
 at a certain  term   $k=K$.
Following  \cite{wat90},  we   ignore all the terms having 
$kh\ge 4(=\xi_\text{max})$, which implies that any term whose  weight $w(kh)$ is
less than  or equal to  
\begin{equation}
w(4)=6.377\cdots \times 10^{-36}   \label{eq-17}
\end{equation}  
is to be omitted. 
If the truncation at other  location $\xi_\text{max}$ is desired, such that  $w(\xi_\text{max})=F$, \cite{wat90}
gives the following approximate formula:
\begin{equation}
\xi_\text{max}\simeq \log \left\{(2/\pi)\log\left[(4/\pi F)\log(2/F)\right]\right\}, \label{eq-18}
\end{equation}
whereas  we have  derived 
\begin{subequations}
\begin{align}
\xi_\text{max}&\simeq\left\{\left[X\left(\log\left[(2/\pi)X\right]-1\right)\right.\right.   \notag \\
&\qquad\qquad\qquad \left.\left.+\log(2/F)\right]\right\}/(X-1),\\
X&=\log\left[(4/\pi F)\log(2/F)\right],
\end{align}  \label{eq-19}
\end{subequations}
\hspace*{-0.1cm}which is somewhat more accurate than Eq.\eqref{eq-18}.
\indent From Eq.\eqref{eq-15}, we have  the following approximation $S_1^{(m)}$ for  $S^{(m)}$: 
\begin{equation}
S_1^{(m)}(\mu)=  
\frac{\pi h}{4}\left\{\sum_{k=K}^0\frac{2-\delta_{0k}}{2}w(kh)Z\bigl[\mu, \phi(kh)\bigr]\right\},       \label{eq-20}
\end{equation}
which requires  evaluations  of the integrand   $Z$ at $(K+1)$ division points. \par
If, on the other hand, an  approximate calculation for $S^{(m)}(\mu)$  is carried out   using    only the values of the integrand  evaluated  at the $K+1$ midpoints    $\xi_{k+\frac{1}{2}}\equiv(k+\frac{1}{2})h ~(k=0, 1, \cdots, K)$ of  the intervals $[\xi_k, \xi_k+h]\quad (k=0, 1, \cdots, K)$, 
another form of approximation  $S_2^{(m)}(\mu)$ results: 
\begin{multline}
S_2^{(m)}(\mu)=
 \displaystyle{\frac{\pi h}{4}\left\{\sum_{k=K}^0\!\!w\bigl[(k+\frac{1}{2})h\bigr] \right. }    \label{eq-21}\\
 \displaystyle{\left.\times Z\bigl[\mu, \phi\bigl((k+\frac{1}{2})h\bigr)\bigr]\right\}. \qquad \qquad }
\end{multline} 
Putting $h^\prime=h/2$, we then have
\begin{multline}
S_3^{(m)}(\mu)=\frac{1}{2}\bigl[S_1^{(m)}(\mu)+S_2^{(m)}(\mu)\bigr]  \\
=\displaystyle{\frac{\pi h^\prime}{4}\left\{\!\sum_{~~k=2K+1}^0\!\!\!\!\frac{2-\delta_{0k}}{2}~w(kh^\prime)Z[\mu, \phi(kh^\prime)]\right\},} \label{eq-22}
\end{multline}
which is a trapezoidal rule   approximation similar to  $S_1^{(m)}(\mu)$ except that the step-size is now $h^\prime$ or $h/2$ instead of $h$.  According to Eq.(28) of \cite{mor90},  the number of significant digits for $S_3^{(m)}$ is expected to be roughly twice as large as  that for $S_2^{(m)}.$ 
\par
The numerical integration for $S^{(m)}(\mu)$ is assumed to have been  accomplished,   
 if the following condition holds:
\begin{equation}
\varepsilon_\text{S}=|S_2^{(m)}(\mu)-S_3^{(m)}(\mu)|/|S_3^{(m)}(\mu)|\le \varepsilon_{\text{S},0}   \label{eq-23}
\end{equation}  
for all the values of $\mu$ that need  to be considered, where $\varepsilon_{\text{S},0}$ is a prescribed small number adopted for  error tolerance~(see Eq.\eqref{eq-31}).
Otherwise,   above procedure is    repeated by
taking this  value of $S_3^{(m)}(\mu)$ as a new one  for  $S_1^{(m)}(\mu)$,  and a combined set of the $2(K+1)$ quadrature points $\phi(\xi_k^\prime)\ (k=0, 1, \cdots, 2K+1)$ with  $\xi_k^\prime=kh^\prime$ and  corresponding weights are assigned to the sets of the quadrature points and the weights for the renewed $S_1^{(m)}(\mu)$ with 
$h^\prime(=h/2)$ being taken as a new value for  the step-size $h$. 
Subsequently, a   set of $2(K+1)$ midpoints 
$\xi_{k+\frac{1}{2}}^\prime\left(\equiv(k+\frac{1}{2})h^\prime\right)$ 
of the intervals $[\xi_k^\prime, \xi_k^\prime+h^\prime]~(k=0, 1,  \cdots, 2K+1)$
 is generated, with which a set of $2(K+1)$ quadrature points  and the weights  can be calculated 
to obtain  an improved  value for  $S_2^{(m)}(\mu)$ in a manner analogous to  Eq.\eqref{eq-21}. \par
Now, let $J$ be  the maximum number of  step-size reductions that we allow. 
Suppose further that the  $S_1^{(m)}(\mu)$  calculation is initiated with a set of  four quadrature points $\phi(kh) \ \ (k=0, 1, 2, 3)$ together with the  step-size  $h=1$.
Then, in  the $j$-th step-size reduction, we  have $h =1/2^{j-1}\ (j=1, 2, \cdots, J)$ and  $h^\prime=1/2^j$. 
The number of midpoints  newly located  is $2^{j+1}$, thereby yielding $K= 2^{j+1}-1$ for Eq.\eqref{eq-21}.
The sum of the number of the existing division points and that of the newly produced  division points is thus  $2(K+1)=2^{j+2}$, which implies $2K+1=2^{j+2}-1$ for Eq.\eqref{eq-22}.
Consequently, the total number  of the quadrature points we need  to calculate   over  the  entire $J$  step-size reductions is   $2^{J+2}$.
All these  quadrature points and the corresponding weights  can be sorted in increasing order of $\xi$ to construct a single numerical table, provided  we assign them  sequential ID numbers $n$ in the following fashion  as   we calculate them:  
\begin{subequations}
\begin{align}
\phi_n&\equiv\phi(\xi_n), \quad w_n\equiv w(\xi_n) \label{eq-24a}\\
\xi_n&= \begin{cases}
                             k   \quad \hspace{0.85cm}(j=0, ~k=0, 1, 2, 3),       \\
                           \displaystyle{ \frac{2k+1}{2^j} }\quad(j\ge 1, ~k=0,  \cdots, 2^{j+1}\!\!\!-1),   \\
              \end{cases}  \label{eq-24b}
\end{align}  
\end{subequations} 
with
\begin{equation}
n=  \begin{cases}
                        2^Jk   \quad \hspace{1.05cm}(j=0, ~k=0, 1, 2, 3),       \\
                        \displaystyle{\frac{2^J(2k+1)}{2^j} }\ \ (j\ge 1, ~k=0,  \cdots, 2^{j+1}\!\!\!-1), 
     \end{cases}  
      \label{eq-25} 
\end{equation}  
where $j=0$ signifies the initial stage prior to entering the step-size reduction procedure for which  $j\ge 1$.
Eq.\eqref{eq-25} shows that $n$ takes $2^{J+2}$ integer values  running  from $0$ through $N$, where $N=2^{J+2}-1$ given by $j=J$ and $k=2^{J+1}-1$.  Consequently, the largest  division point corresponds to  $\xi_N=Nh^\prime=2^2-2^{-J}(<4)$.
Conversely, Eq.\eqref{eq-25} permits  us to extract from the above-mentioned  table the relevant set of quadrature points $\phi_n$'s and their  weights $w_n$'s  required to calculate  the value of  $S_2^{(m)}$
necessary  to obtain that of $S_3^{(m)}$ for a given value of  $j$.  This   facilitates an automatic accuracy adjustment  in  solving Eq.\eqref{eq-6} for $H^{(m)}(\varpi_0, \mu)$ by means of a successive approximation  using  the DE-formula.  
%%%%%%%%%%%%%%%%%%%%%%%%%%%%%%%%%%%%%%%%%%%%%%%%%%%%%%%%%%%%%
\subsection{Iterative Scheme to Solve for $H$-Functions}
%%%%%%%%%%%%%%%%%%%%%%%%%%%%%%%%%%%%%%%%%%%%%%%%%%%%%%%%%%%%%
For iterative solution of Eq.\eqref{eq-6}, we employ the following  simultaneous set of algebraic equations: \\
 \vspace*{-0.5cm}
\begin{multline}
H^{(m)}(\varpi_0,\mu_n^\pm)_\text{new}=\bigg\{\!\!\sqrt{1-2\psi_0^{(m)}} 
+S^{(m)}(\mu_n^\pm)_\text{old} \biggr\}^{-1}  \\
\quad  (n=0, 1, \cdots, N+1), \label{eq-26}
\end{multline} 
where $N+1=2^{J+2}$  as indicated in \S2.2.
Here, $H^{(m)}(\varpi_0, \mu_n^\pm$ $)_\text{new}$ are  updated   values of  $H^{(m)}(\varpi_0,$
$\mu_n^\pm)$,  while $S^{(m)}(\mu_n^\pm)_\text{old}$ are  those of $S^{(m)}(\mu_n^\pm)$ evaluated with the  set of  $H^{(m)}(\varpi_0, \mu_k^\pm)_\text{old}$ \ \ $(k=0, 1, \cdots, N+1)$ that we wish to improve, and 
\begin{equation}
\mu_n^\pm=\frac{1\pm \phi_n}{2} \quad (n=0, 1, \cdots, N), \quad \mu_{N+1}^\pm=0, \label{eq-27}
\end{equation}
where, as mentioned before,  $\mu_n^+$ corresponds to the plus sign  and $\mu_n^-$  to the minus sign on the right-hand side of   Eq.\eqref{eq-27} respectively, with special cases that $\mu_0^+=\mu_0^-=\frac{1}{2}$ and that $\mu_{N+1}^-=\mu_{N+1}^+=0$.
The  non-quadrature points $\mu_{N+1}^\pm$ are explicitly included  in the above iterative procedure for the reason that 
the value of $H^{(m)}(\varpi_0, 0)_\text{new}$ is required  in order to normalize   $H^{(m)}(\varpi_0, \mu_n^\pm)_\text{new}\quad (n=0, 1, \cdots, N)$~(see Eq.\eqref{eq-30} below) to create a new starting set of $H^{(m)}(\varpi_0, \mu_n^\pm)_\text{old}$'s, if another iteration needs to be performed.\par 
For a given value of $\varpi_0$, a starting  approximation for $H^{(m)}(\varpi_0, \mu_n^\pm)_\text{old}$ required to evaluate the right-hand side of Eq.\eqref{eq-26}  is created  in the manner  proposed  by \cite{kaw15}:
\begin{enumerate}
\item[(i)] $m=0$: the approximate formula of Kawabata and Limaye (2011, 2013)
%\cite{kaw11, kaw13}
 developed for the isotropic scattering $H$-function is used.  It should be reminded that the maximum relative error of this formula is claimed to be  $2.1\times 10^{-4}~\%$.
\item[(ii)] $m\ge 1$:  the solution for  $H^{(m-1)}(\varpi_0, \mu_n^\pm)$
 is substituted for $H^{(m)}(\varpi_0, \mu_n^\pm)$.
\end{enumerate}
The relative deviations  $\varepsilon_\text{H}(\varpi_0, \mu_n^\pm)$ of $H^{(m)}(\varpi_0, \mu_n^\pm)_\text{new}$ are then calculated:  
\begin{multline}
\varepsilon_\text{H}(\varpi_0, \mu_n^\pm) \\
=\left|\left(H^{(m)}(\varpi_0,\mu_n^\pm)_\text{old}/H^{(m)}(\varpi_0, \mu_n^\pm)_\text{new}\right)-1\right|  \\  
(n=0, 1, \cdots, N+1), \label{eq-28}
\end{multline}
and the iteration is  terminated if  the condition
\begin{equation}
\varepsilon_\text{H}^\text{max}\equiv\max\{\varepsilon_\text{H}(\varpi_0, \mu_n^\pm)\}\le \varepsilon_{\text{H},0}(=10^{-15})  \label{eq-29}
\end{equation}
is satisfied at  all of $\mu_n^\pm$  for a given value of   $\varpi_0$.
The resulting $N+2$ values of  $H^{(m)}(\varpi_0,\mu_n^\pm)_\text{new}\quad (n=0, 1, \cdots, N+1)$ are
adopted as a base set of the  desired  solution.
Otherwise, we proceed to the next round of iteration employing  $H^{(m)}(\varpi_0,$ $\mu_n^\pm)_\text{new}\quad (n=0, 1, \cdots, N+1)$ obtained above as  a new set of approximate  values  for  $H^{(m)}(\varpi_0, \mu_n^\pm)_\text{old}\quad (n=0, 1, \cdots, N+1)$  after  applying 
 the  normalization procedure  adopted by  \cite{bos83}:
\begin{multline}
H^{(m)}(\varpi_0, \mu_n^\pm)_\text{old}\\
=H^{(m)}(\varpi_0, \mu_n^\pm)_\text{new}/H^{(m)}(\varpi_0, 0)_\text{new}. \label{eq-30}
\end{multline}
together with the obvious condition $H^{(m)}(\varpi_0, 0)_\text{old}=1$.
According to \cite{kaw15}, this step is crucial in order to secure the convergence to the solution.
With  the base  set  of  converged values for $H^{(m)}(\varpi_0, \mu_n^\pm)$\quad$(n=0, 1, \cdots, N+1)$ available, we can calculate  without recourse to any interpolation procedure the values of $H^{(m)}(\varpi_0, \mu)$ for  arbitrary values of $\mu$  using Eq.\eqref{eq-6} with  the  same value of $\varpi_0$ employed above,  although we have to repeat  calculations of $S^{(m)}(\mu)$  again applying  the DE-formula  with the same convergence criterion indicated by  Eq.\eqref{eq-23}. 
%%%%%%%%%%%%%%%%%%%%%%%%%%%%%%%%%%%%%%%%%%%%%%%%%%%%%%%%%%%%
\section{Numerical Calculations and Results}
%%%%%%%%%%%%%%%%%%%%%%%%%%%%%%%%%%%%%%%%%%%%%%%%%%%%%%%%%%%%
All of our calculations for the present work are carried out in double-precision arithmetic.
Using  Eqs.\eqref{eq-6}, \eqref{eq-23},  \eqref{eq-26}, and \eqref{eq-28},  we realize  
\begin{align}
\varepsilon_\text{H}&=\frac{|H_\text{old}^{(m)}-H_\text{new}^{(m)}|}{H_\text{new}^{(m)}} \le H_\text{old}^{(m)}S_3^{(m)}\varepsilon_{\text{S},0} \simeq H_\text{new}^{(m)}S_3^{(m)}\varepsilon_{\text{S},0}\notag \\
&=\frac{S_3^{(m)}}{\sqrt{1-\psi_0^{(m)}}+S_3^{(m)}}\varepsilon_{\text{S},0} \le \varepsilon_{\text{S},0}.  \label{eq-31}
\end{align} 
Letting  $\varepsilon_{\text{S },0}=\varepsilon_{\text{H},0}(=10^{-15})$, we  automatically satisfy Eq.\eqref{eq-29}, the convergence criterion for the function $H^{(m)}$, 
whenever  Eq.\eqref{eq-23} holds for the integral term $S^{(m)}$.
 Here, it must be mentioned that  the magnitudes of the absolute  errors involved in resulting  values of $S_3^{(m)}$ due to the discretization of the integral~(see Eq.\eqref{eq-22}) are  
 of the order of  $(S_2^{(m)}(\mu)-S_3^{(m)}(\mu))^2$ or
$S_3^{(m)}(\mu)^2\varepsilon_\text{S}^2$  according to \cite{tm74}.
 Furthermore,   Eqs.\eqref{eq-7b} and \eqref{eq-7c} of the present work coupled  with Eq.(7) of Chap.V of \cite{cha50} indicate  $S_3^{(m)}(\mu)\le 1$. 
Therefore, the magnitude of the discretization error of $S_3^{(m)}$ should be around $\varepsilon_\text{S}^2$, which is $10^{-30}$, if  $\varepsilon_{\text{S},0}=10^{-15}$ is  adopted.  
In other words, the absolute errors associated with  the values of  $S_3^{(m)}(\mu)$ would 
primarily be due to the round-off errors which are of the order of 
 $10^{-16}$ in the case of double-precision  calculations.  This in turn   implies that the absolute errors involved in the values of $H^{(m)}$ produced  by Eq.\eqref{eq-26} would be  determined also by the round-off errors 
in the values of $(1-\psi_0^{(m)})^{1/2}+S^{(m)}$. \par
An attractive feature of the DE-formula is that it  enables us to efficiently perform an automatic step-size adjustment to produce  results of numerical integrations with  considerably high   accuracy. 
In order to implement such  procedure in the numerical integration for $S^{(m)}$,  we have varied the value of $J$, the maximum number to be allowed for step-size reductions, from 4 to 7, to find $J=6$ is  optimum. This  choice yields   $N=255$ according to Eq.\eqref{eq-25}, so that for  a given value of $\varpi_0$ and for a specific value of $m$, we  need to solve Eq .\eqref{eq-26} for $H^{(m)}(\varpi_0, \mu_n^\pm)$  at  514  division points
 $\mu_n^\pm\quad (n=0, 1, \cdots, 255, 256)$ taking into account  the fact that   $H^{(m)}(\varpi_0, \mu_0^-)=H^{(m)}(\varpi_0, \mu_0^+)$ and that $H^{(m)}(\varpi_0, \mu_{N+1}^+)=H^{(m)}(\varpi_0, \mu_{N+1}^-)$. \par
As for the values of the single scattering albedo $\varpi_0$ to be used for tabulations of the final results  for $H^{(m)}(\varpi_0, \mu)$, the same 56 values as those employed by \cite{kaw16}
are used: 
\begin{align}
\varpi_0=&10^{-3}, 0.1(0.1)0.5, 0.55(0.05)0.75, 0.8(0.02)0.9, \notag\\
&0.91(0.01)0.95, 0.96(0.005)0.98, \notag\\
&0.982(0.002)0.99, 0.991(0.001)0.998, \notag\\
& 0.9985(0.0005)0.9995, 0.9996(0.0001)0.9999, \notag\\
&1-10^{-k}\  (k=5, 7, 9, 10(1)14), 1,\hfill \label{eq-32}
\end{align}
where a parenthesized number in between a pair of two figures indicates the increment to be successively added to a preceding figure to get the next one, such that $0.1(0.1)0.5$ means a set of $0.1, 0.2, 0.3, 0.4$, and $0.5$.\par
In the course of the experiments, we were also made aware  of the importance of inspecting  
   numerical accuracies  of the resulting solutions  
  at  $\mu$'s   much smaller than 0.05 as well, and hence   14 new values are presently 
   added to the set of 22  previously employed   by \cite{kaw15, kaw16}, to have  
  the   following set of 36 $\mu$-arguments for tabulations:
\begin{align}
\mu=&0, 10^{-k}\ (k=12(-1)6), 5\times 10^{-6}, 10^{-5}, 5\times 10^{-5},\notag\\
&10^{-4}, 5\times 10^{-4}, 10^{-3}, 5\times 10^{-3}, 10^{-2}, 5\times 10^{-2},\notag\\
&0.1+ 0.05~(k-1)~~(k=1,2, \cdots, 19). \hfill  \label{eq-33} 
\end{align}
\indent As indicated in Table 1, we carry out numerical calculations of the $H$-functions for  27 cases of phase functions $P(\Theta)$ taken  from those employed by \cite{kaw15} setting aside  those which  exhibit negative
values in  certain regions of scattering angle $\Theta$. 
Here,  $M$ shown  in the second column  indicates the highest degree of the Legendre polynomial to be retained in Eq.\eqref{eq-8}, and $x_m$'s in  columns 3 through 5 are the required expansion coefficients as has already been mentioned.  \par 
The sixth  column shows  the ID numbers assigned in \cite{kaw15}, and each row of the last column  gives  an abbreviated name of  the relevant scattering law or  references: 
  'ISO' for No.1 row designates  the  isotropic scattering,   'LIN' for No. 2 through No. 7 is for   linearly anisotropic scattering (the two-term phase function),  and  No.8 through No.17 are examples of the three-term phase function, and, in particular,  'RAY' for No.8 designates   the Rayleigh scattering.  They are  important for the present work  in that some sample numerical values of high accuracy are  available for comparison through the works of \cite{vii86},  \cite{jab15}, and \cite{kaw16}. 
 The rows No.18 through No.27 are samples for  the full four-term phase function ($M=3$), taken from \cite{kol72} and \cite{hul80},  each of which gives rise to   four Fourier components  $\psi^{(m)}(\mu) \quad (m=0, 1, 2, 3)$ of the corresponding characteristic function.   \par
  We solve Eqs.(\ref{eq-26})  iteratively for $H^{(m)}(\varpi_0,\mu_n^\pm)$ corresponding to 83 Fourier components of  characteristic functions originating from the 27 phase functions.
It must be stressed here that for the convenience of  comparison, we have repeated  the $H$-function calculations of \cite{kaw15} using the  new set of $\mu$-arguments shown by Eq.\eqref{eq-33}
and the 256-point rather than 128-point Gauss-Legendre quadrature. 
As for the convergence criteria Eqs.\eqref{eq-23} and \eqref{eq-29}, we employ    $\varepsilon_{\text{S},0}=\varepsilon_{\text{H},0}=10^{-15}$ as has already been discussed.  \par
Our final  results for  the  sets of $H^{(m)}(\varpi_0, \mu_n^\pm)\quad (n=0, 1, \cdots, 256)$ employing  the  27  phase functions show that  the maximum  of all the values for  $\varepsilon_\text{H}^\text{max}$  is  $1.0\times 10^{-15}$(see Eq.\eqref{eq-29}), 
taking place in  the case with $m=0$,  $\varpi_0=0.86$, and 
$\mu_{57}^+(=0.9601830265159764)$ for  the phase function No.13, where 
$H^{(0)}(0.86, \mu_{57}^+)_\text{new}=1.776790015838130$.
On the other hand, the maximum relative deviation $\max|H^{(m)}(\varpi_0, \mu)_\text{old}/H^{(m)}(\varpi_0,\mu)_\text{new}$ $ -1|$ of the entire tabular values of  $H^{(m)}(\varpi_0, \mu)_\text{new}$,  where those for $\mu$ are given by  Eq.\eqref{eq-33}, is $9.7\times 10^{-16}$ found for  $H^{(0)}(0.994, 0.05)_\text{new}$  $=1.146211415422285$ obtained with  the phase function No. 11.\par
Table 2 presents the values of the $H$-function for conservative isotropic scattering obtained using  three types of computational techniques: (a) $H(\varpi_0, \mu)_\text{DE}$ shows our  present results based on the iterative solution applying the DE-formula, (b)$H(\varpi_0, \mu)_\text{Gauss}$ gives those found by the iterative scheme of \cite{kaw15} with the 256-point Gauss-Legendre quadrature, and (c)$H(\varpi_0, \mu)_\text{Ana}$ 
presents, as a reference of comparison,  the results  obtained  using the analytical solution with numerical integration performed with  the new set of $\mu$ values by means of the DE-formula as in  \cite{kaw16}.  As has already been pointed out,  the results for  (c) agree with those of \cite{jab15} within one unit difference in  the 15-th decimal place. \par
The fifth and sixth  columns show respectively the deviations of (a) and (b) from (c) multiplied by a factor $10^{15}$, viz., 
\begin{subequations}
\begin{align}
\Delta_\text{DE}\hspace{0.3cm}&=10^{15}[H(1, \mu)_\text{DE}-H(1, \mu)_\text{Ana}], \label{eq-34a}\\
\Delta_\text{Gauss}&=10^{15}[H(1, \mu)_\text{Gauss}-H(1, \mu)_\text{Ana}].  \label{eq-34b}
\end{align} \label{eq-34}
\end{subequations}
It can be seen that the results for (a) and (c) are in close agreement with each other with differences $\Delta_\text{DE}$ of no more than  one unit in the 15-th decimal place even if   we have included 14 considerably  small values for $\mu$-arguments. 
On the other hand, the results for (b) exhibit significantly larger deviations from those of (c) in the domain of   $\mu<10^{-3}$.
In fact, we notice  differences by  one unit even  in the 6-th decimal place at $\mu=5\times 10^{-6}$ and
 $10^{-5}$ despite the fact that we have doubled the number of the Gaussian points  for numerical integrations in (b) in comparison with that employed by \cite{kaw15}.  Obviously,  we also have to carefully watch  accuracies of  numerical results obtained for $\mu$ much less than $0.05$, the smallest argument value often employed for tabulations by various investigators. \par
 It should  be noted that, as a supplemental check, we have  made a comparison of our  values of $H(0.5, \mu)_\text{DE}$ and $H(1, \mu)_\text{DE}$
  rounded to the 13-th decimal place with those of \cite{vii86}, to find one unit differences in the 13-th decimal place  at $\mu=0.2, 0.4, 0.6$, and $0.7$ for the former, but  no difference at all  for the latter. 
  All these comparisons  give much 
  credence to our present numerical scheme.  \par
The bottom  five rows of Table 2  show the  values of the 0-th trough 4-th order moments  of the $H$-function calculated for the conservative isotropic scattering together with 
the deviations  (a) $\Delta_\text{DE}$ and (b) $\Delta_\text{Gauss}$:
\begin{subequations}
\begin{align}
\Delta_\text{DE}\hspace{0.3cm}&=10^{15}\left[\alpha_\text{$k$, DE}-\alpha_\text{$k$, Ana}\right], \label{eq-35a}\\
\Delta_\text{Gauss}&=10^{15}\left[\alpha_\text{$k$, Gauss}-\alpha_\text{$k$, Ana}\right], \label{eq-35b}\\
& \hspace{3cm}(k=0, 1, 2, 3, 4).  \notag
\end{align}  \label{eq-35}
\end{subequations} 
In the case of the moment calculations, both (a) $\alpha_{k, \text{DE}}$ and (b) $\alpha_{k, \text{Gauss}}\quad (k=0, 1, 2, 3, 4)$ are in close agreement with those given by(c)  $\alpha_{k, \text{Ana}}$, although the  results for (a) appear to be slightly  more superior to those of (b).  
It should also be noted  that  the values of  $\alpha_{0, \text{DE}}$ obtained for  the entire 56 cases  of $\varpi_0$~(Eq.\eqref{eq-32}) agree with the following exact values~\citep[see, e.g.,][]{cha50, iva73, hul80} within an absolute error of $4.44\times 10^{-16}$:
\begin{equation}
\alpha_0=(2/\varpi_0)\big[1-\sqrt{1-\varpi_0}\big].  \label{eq-36}
\end{equation} 
For additional check, a comparison has been made between  our values of
 $\alpha_{k, \text{DE}}$ 
 and
  $\alpha_{k,\text{Gauss}}\quad (k=0, 1, 2, 3, 4)$ rounded to the 13-th decimal place and those of \cite{vii86} calculated for $\varpi_0=0.5$ and $1$, to find that they are all  in complete  agreement except that $\alpha_{1,\text{DE}}$ and  $\alpha_{1, \text{Gauss}}$ for $\varpi_0=1$  differ from those  of \cite{vii86} by one unit  in the 13-th decimal place
due possibly  to round-off errors.  
On the other hand, the values given to the 15-th decimal place by \cite{jab15} in his Appendix for $\alpha_k \quad (k=0, 1, 2)$ of  the conservative scattering case
are $2.000000000028379$, $1.154700538378986$, and $0.820352482149141$, respectively, whose deviations from ours are correspondingly $28379\times 10^{-15}$, $-265\times 10^{-15}$, and $15\times 10^{-15}$. \par 
Table 3 shows the  values of  $H^{(m)}(1, \mu)_\text{DE}$\quad $(m=0, 1, 2)$ calculated for  conservative Rayleigh scattering  whose phase function is characterized by the expansion coefficients  given in the No.8 row of Table 1.   They are found to agree almost perfectly with those of \cite{vii86}, if rounded  to the 13-th decimal place, with
 differences by  one unit  in the last significant digit taking place only in  $H^{(2)}(0.5, 0.6)$, $H^{(1)}(1, 0.6)$, and $H^{(2)}(1, 0.8)$.
 Also shown in  columns 5 through 7 are the deviations
 of those of $H^{(m)}(1, \mu)_\text{Gauss}$\quad $(m=0, 1, 2)$~(their values not reproduced here for the lack of space) recomputed using the procedure of \cite{kaw15} but for 36 $\mu$-values enumerated by Eq.\eqref{eq-33} :
\begin{align}
\Delta_\text{Gauss}^{(m)}&=10^{15}\left[H^{(m)}(1, \mu)_\text{Gauss}-H^{(m)}(1, \mu)_\text{DE}\right] \notag \\
&\hspace{3.5cm}(m=0, 1, 2). \label{eq-37}
\end{align}
We again notice that there occur considerable degrees of deviations of the results in the domain  
 of  $\mu$  less than  $0.05$ in the cases of  $m=0$ and $2$, although  no  such deviations
are found  for  $m=1$.   
For $\mu \ge 0.05$,  on the other hand, only the values of $H^{(0)}(1, \mu)_\text{Gauss}$ show  comparatively large  deviations. They are nevertheless
 less than two units in the 14-th decimal place.  \par
As in Table 2, the columns 2 through 4 in the bottom five rows give  the resulting values of    $\alpha_{k, \text{DE}}^{(m)}\quad (k=0, 1, 2, 3, 4; m= , 1, 2)$, whereas the columns 5 through 7 show 
deviations of  the values of  $\alpha_{k, \text{Gauss}}^{(m)}\quad (k=0, 1, 2, 3, 4; m=0, 1, 2)$ with respect to  those obtained in the present work with the DE-formula:
\begin{align}
\Delta_\text{Gauss}^{(m)}&=10^{15}\left[\alpha_{k, \text{Gauss}}^{(m)}-\alpha_{k, \text{DE}}^{(m)}\right]\quad
 (m=0, 1, 2).  \label{eq-38}
\end{align}
The values of the moments calculated by two schemes are obviously in close agreement with differences being at most 8 units in the 15-th decimal place. Furthermore, they agree with those obtained by \cite{vii86} for  $\alpha_k^{(m)}\ \ (k=0, 1, 2; m=0, 1, 2)$ to the 13-th decimal place with one unit difference in the last decimal found only for  $\alpha_1^{(0)}$. \par
Columns 2 through 5 of Table 4  show,  as in Table 4 of \cite{kaw15}, our new   results  $H^{(m)}(1, \mu)_\text{DE}\quad (m=0, 1, 2, 3)$  for the conservative scattering  arising from  the phase function No. 25 (No.34 in \cite{kaw15})  displayed in Fig. 1, which yields four  Fourier components  $\psi^{(m)}(\mu)$ $(m=0, 1, 2, 3)$ for  the corresponding  characteristic function  as  plotted in Fig.2.
Our values for $H^{(0)}(1,\mu)$ rounded to the third decimal place are found to agree  with those  of \cite{kol72}  with a difference by one unit  in the last digit occurring  only at $\mu=0.6$ \citep[no numerical results are given  for $m\ge 1$ in ][]{kol72}.
 The 6-th column of Table 4 gives the deviations  $\Delta_\text{Gauss}^{(0)}$(see Eq.\eqref{eq-37}) calculated for  the values of $H^{(0)}(\varpi_0, \mu)_\text{Gauss}$ in 
 comparison with  the present values $H^{(0)}(1, \mu)_\text{DE}$.
Clearly, they  are again   considerably   large for $\mu<0.05$  as we have observed in Tables 2 and 3.
A similar conclusion applies also to the results (not shown here) for  $m=1$ through 3.
Yet,  the values  of $H^{(0)}(\varpi_0,\mu)_\text{Gauss}$  for  $\mu\ge 0.05$  still remain sufficiently accurate. In fact, the present calculations rounded to the 10-th decimal place are in complete agreement 
with  those  of \cite{kaw15} in this range of $\mu$.\par
Columns 2 through 5 of the  bottom five rows  give the present values of the moments $\alpha^{(m)}_{k, \text{DE}}\quad (k=0, 1, 2, 3, 4; m=0, 1, 2, 3)$, and the column 6 shows the deviations $\Delta_\text{Gauss}^{(0)}$  defined by Eq.\eqref{eq-38} for $\alpha_{0,\text{Gauss}}^{(0)}$.  Interestingly,  the values of the moments  produced by the two independent  procedures are in close agreement in spite of the comparatively  huge  differences noticeable in  the values of  $H^{(0)}(\varpi_0, \mu)$ for $\mu < 0.05$.
%%%%%%%%%%%%%%%%%%%%%%%%%%%%%%%%%%%%%%%%%%%%%%%%%%%%%%%%%%%%%
\section{Conclusions}
%%%%%%%%%%%%%%%%%%%%%%%%%%%%%%%%%%%%%%%%%%%%%%%%%%%%%%%%%%%%%
Accurate numerical evaluations of the Ambartsumian-Chandrasekhar $H$-functions are important for various applications, but actual calculations are rather challenging even in the case of  isotropic scattering  and require very careful treatments to avoid introducing errors as has already been pointed out by \cite{jab15}\citep[see also][]{das08}. \par
For this reason, we have developed a straightforward  iterative scheme   to solve 
Eq.\eqref{eq-6}, a variant of the Ambartsumian-Chandrasekhar equation for the $H$-functions corresponding to  the so-called four-term phase function, viz., the phase function expressible by retaining the first four terms or less in its  Legendre polynomial series expansion indicated by Eq\eqref{eq-8}: we have thereby taken  advantage of the superior nature  of the double-exponential formula (DE-formula) of \cite{tm74} for numerical integrations.  
The absolute errors involved in resulting  numerical values for the Fourier components of the $H$-function 
 are  supposed  to be much less than $10^{-14}$,  and hence the solutions should be   accurate to 15 significant figures. The numerical results presented  in Tables 2, 3, and 4 of the present  work must therefore be useful as benchmarks.\par
Although our numerical checks  have been restricted only  to the 27 selected cases of fairly simple analytical phase functions,  the method is likely to prove  valid also for more complex phase functions  so  long as their characteristic functions are analytic over the   $\mu$-interval $[0, 1]$, or  have singularities only at edge points due to the fact that the DE-formula is known to be especially suited under  such circumstances\citep{tm74}. The superiority of the DE-formula over the popular Gauss-Legendre quadrature is  quite evident  
 in comparison of the present results with  those obtained by  \cite{kaw15} who employed the Gauss-Legendre quadrature.\par
 Unlike  the Gauss-Legendre quadrature, the division points and the associated weights of the DE-formula can be easily  calculated  for any  degree desired,
 so that upgrading  numerical accuracy of the formula is of no difficulty.
 In addition, all the values that have been obtained for  relevant integrands  up to a certain stage of approximation 
  can be  utilized with zero waste to further improve  the degree of approximation.
This characteristic feature makes it especially simple to implement an automatic error-control capability in the iterative scheme we employ as has been done in the present work.    \par
 One important lesson learned in this  work is that we must also
be concerned with  the  numerical accuracy of  the $H$-functions 
  for $\mu$-values much less than 0.05,
 the domain often  neglected  as is  the case with  \cite{kaw15}.  This is 
  why a set of  14  $\mu$-values were newly added  during the course of  the present work to the 22  adopted by \cite{kaw15} for tabulations. \par
Furthermore, we have had an excellent opportunity to assess the maximum relative error expected for the rational approximation formula of Kawabata and Limaye (2011) (see Kawabata and Limaye 2013 for Erratum)
%\cite{kaw11}\citep[see][for Erratum]{kaw13}
 for the $H$-function for isotropic scattering. 
 To do so, 10 extra values for $\varpi_0$, viz., $10^{-10}$, $\sqrt{10}\times 10^{-10}, 10^{-9}, 10^{-8}, \sqrt{10}\times 10^{-8}, 10^{-7}, \sqrt{10}\times 10^{-7}, 10^{-6}, 10^{-5}$, 
 and $10^{-4}$, were added on the basis of \cite{jab15} to those indicated by Eq.\eqref{eq-32}, and  the isotropic scattering $H$-function was then evaluated by the present method as well as by using  the Kawabata-Limaye formula for $66\times 36$ combinations of $(\varpi_0,\mu)$, to find the maximum relative error of the formula is $2.4\times 10^{-6}$, viz.,  $2.4\times 10^{-4}$\% as opposed to $2.1\times 10^{-4}$\% noted   by \cite{kaw11}.\par
According to  \cite{hul80},  the  $H$-function method  becomes impracticable as a means  to directly calculate the reflection functions of semi-infinite, vertically homogeneous  atmospheres characterized by phase functions more complex than those we have considered in this work.
This implies that for  realistic phase functions,  solving the Ambartsumian equation for reflection functions
or alternatively making use of the analytical representations  derived by \cite{rog16} for  reflection function, plane and spherical albedos is likely to be more efficient as has already been mentioned.
In view of this and as an application of the isotropic scattering $H$-function, an attempt has therefore been made  to solve by means of a successive approximation the Ambartsumian equation for a semi-infinite, vertically homogeneous  atmosphere whose scattering law is specified  by either the single-term  or the two-term Henyey-Greenstein phase function: the values of the relevant  parameters  are  taken identical to those employed by \cite{rog16} to produce their Tables 1 and 2 and Figures 5 add 6.
The details of our computational procedure and  the numerical results  are summarized  in Appendix. 
Primitive though it may seem,  our  procedure is found to be sufficiently competitive,  closely simulating   their results.
Yet  a need for  implementing a faster  algorithm in our procedure still remains   in order to  
 deal with problems involving highly anisotropic phase functions in  near-conservative scattering.
  \par

%% Included in this acknowledgments section are examples of the
%% AASTeX hypertext markup commands. Use \url without the optional [HREF]
%% argument when you want to print the url directly in the text. Otherwise,
%% use either \url or \anchor, with the HREF as the first argument and the
%% text to be printed in the second.
%%%%%%%%%%%%%%%%%%%%%%%%%%%%%%%%%%%%%%%%%%%%%%
%%%%%%%%%%%%%%%%%%%%%%%%%%%%%%%%%%%%%%%%%%%%
%%%%%%%%%%%%%%%%%%%%%%%%%%%%%%%%%%%%%%%%
\acknowledgments
The author is grateful to the anonymous referee for his or her constructive and highly enlightening comments. Thanks are also due to A. Jablonski for communicating his recent work  on numerical evaluation of the $H$-function for isotropic scattering.\par
This is a pre-print of an article published in Astrophysics and Space Science, January 2018, 363:1. The final authenticated version is available on line at:\\ https://doi.org/10.1007/s10509-017-3218-5.\\ (First Online: 01 December 2017)
%%%%%%%%%%%%%%%%%%%%%%%%%%%%%%%%%%%%%%%%
%%%%%%%%%%%%%%%%%%%%%%%%%%%%%%%%%%%%%%%%%%%
%%%%%%%%%%%%%%%%%%%%%%%%%%%%%%%%%%%%%%%%%%%%%%
\appendix
\twocolumn
\setcounter{table}{1}
\section{Numerical Calculations of Plane and Spherical Albedos for Semi-Infinite, Vertically Homogeneous Media}
The Ambartsumian equation to determine  the $m$-th order Fourier coefficient of  the  reflection function $R^{(m)}(\mu, \mu_0)~(\mu, \mu_0\in [0,1])$  for  a semi-infinite medium takes  the  following form\citep{sob75,yan97,mis99}:
\begin{multline}
R^{(m)}(\mu, \mu_0)=\frac{1}{4(\mu+\mu_0)}\left\{
P^{(m)}(-\mu, \mu_0) \phantom{\int_0^1}\right. \\
+2\mu\int_0^1R^{(m)}(\mu,\mu^\prime)P^{(m)}(\mu^\prime,\mu_0)d\mu^\prime\\
+2\mu_0\int_0^1P^{(m)}(\mu,\mu^\prime)R^{(m)}(\mu^\prime,\mu_0)d\mu^\prime  \\
\left.+4\mu\mu_0\int_0^1R^{(m)}(\mu, \mu^\prime)\left[\int_0^1 \right.   \right.
            P^{(m)}(-\mu^\prime,\mu^{\prime\prime})    \\
\left.  \left.  \phantom{\frac{1}{2}\int_0^1}  \times R^{(m)}(\mu^{\prime\prime},\mu_0)d\mu^{\prime\prime}\right]d\mu^\prime\right\},  \label{eq-A1}
\end{multline}
where $P^{(m)}(\mu, \mu_0)$ is the $m$-th order Fourier coefficient of the phase function of our interest~(see Eq.(A2) below).
It must be stressed that  $P^{(m)}$  includes the single scattering albedo $\varpi_0$ as a
multiplicative factor~(Eq.\eqref{eq-8} of the main text). 
Eq.\eqref{eq-A1} is usually discretized by approximating  the integrals with respect $\mu^\prime$ and $\mu^{\prime\prime}$ by  an $N_\mu$-th order quadrature  as shown  by Eq.(29) of \cite{mis99},  yielding a system  of  $N_\mu\times N_\mu$ simultaneous equations, which we intend to  solve  here by a  successive approximation method  to investigate how closely we can reproduce the results of Tables 1 and 2 as well as those in Figs. 5 and 6 of \cite{rog16}. \par
For arbitrary phase functions, the  Fourier coefficients $P^{(m)}$ involved in Eq.\eqref{eq-A1} can be numerically evaluated  by  
\begin{multline}
P^{(m)}(u, \mu_0)=\frac{1}{\pi}\int_0^\pi\!\!P(\Theta)\cos m\phi^\prime d\phi^\prime\\
\simeq \frac{1}{\pi}\sum_{n=1}^{N_\phi} w_nP\Large(\cos^{-1}[u\mu_0+\sqrt{(1-u^2)(1-\mu_0^2)} \cos \phi_n]\Large)\\
    \phantom{\frac{1}{AAA}}  \times\cos(m\phi_n),   \label{eq-A2}
\end{multline}
where $u$ designates either  $\mu$ or $-\mu$, while 
$\phi_n~(\in [0, \pi])$ and $w_n$ are the $n$-th division point and the associated integration weight of an  $N_\phi$-th order numerical quadrature employed. \par
However, in the case of  the (single-term) Henyey-Greenstein phase function having an anisotropy parameter $g~\in (-1, 1)$:   
\begin{equation}
P(\Theta)=P_\text{HG}(\Theta; g)=\frac{\varpi_0~(1-g^2)}{(1+g^2-2g\cos \Theta)^{3/2}},    \label{eq-A3}
\end{equation}
the  azimuth angle-averaged term  $P^{(0)}(u, \mu_0)$ can be expressed 
 in the following manner 
 \citep[][on p.333]{hul80}:
\begin{align}
P^{(0)}(u, \mu_0)&=\varpi_0[(1-g^2)/\sqrt{\alpha+\beta}(\alpha-\beta)] \notag\\ 
& \times (2/\pi)E(\pi/2, \sqrt{2\beta/(\alpha+\beta)}~),    \label{eq-A4}
\end{align} 
where $E(\pi/2, k)$ is the complete elliptic  integral of the second kind, while $\alpha$ and $\beta$ are defined as 
\begin{subequations}
\begin{align}
\alpha&=1+g^2-2gu\mu_0,      \label{eq-A5a} \\
\beta&=2|g|\sqrt{(1-u^2)(1-\mu_0^2)}.     \label{eq-A5b}
\end{align}
\end{subequations}
Because of the symmetry relations present in  phase functions~\citep{han74}, we only need to calculate  $P^{(m)}(-\mu, \mu_0)$ and $P^{(m)}(\mu, \mu_0)$ to solve Eq.\eqref{eq-A1}.
However,  in applying a numerical quadrature to the integrals  in Eq.\eqref{eq-A1}, it is crucial to make sure the normalization condition \citep{hov04}
\begin{multline}
\left|\frac{1}{2\varpi_0}\sum_{n=1}^{N_\mu}
\left[P^{(0)}(-\mu_n,\mu_k)+P^{(0)}(\mu_n,\mu_k)\right]w_n-1\right| \\
\equiv B(\mu_k)\le \varepsilon_\text{norm},\quad \quad (k=1, 2, \cdots, N_\mu),                   \label{eq-A6}
\end{multline} 
is satisfied  by the $N_\mu$-th order quadrature to avoid causing an artificial absorption, where $\varepsilon_\text{norm}$ is a prescribed numerical value for error tolerance, while  $\mu_k$ and $\mu_n$ signify the quadrature points.
Whether or not a chosen value for $N_\mu$ is adequate can be assessed to a large extent by inspecting $\max B(\mu_k)\ \ (k=1,2,\cdots, N_\mu)$.
We  iteratively achieve this renormalization  with   
 $\varepsilon_\text{norm}=10^{-14}$ following  the procedure of  \cite{han71}.\par
To set up  a starting approximation for $R^{(m)}(\mu, \mu_0)$ on the right-hand side of Eq.\eqref{eq-A1}, we treat both  single scattering and  second-order scattering rigorously, but  approximate 
all the  higher order  scatterings by  isotropic scattering. Then for $m=0$, we have 
\begin{multline}
R^{(0)}(\mu, \mu_0)=\frac{1}{4(\mu+\mu_0)}\left\{P^{(0)}(-\mu,\mu_0)\right.\\
\left.+\frac{\mu_0}{2}\int_0^1\!\!P^{(0)}(\mu,\mu^\prime)P^{(0)}(-\mu^\prime, \mu_0)d\mu^\prime/(\mu^\prime+\mu_0)\right.                                        \\
 +\frac{\mu}{2}\int_0^1\!\!P^{(0)}(-\mu, \mu^\prime)P^{(0)}(\mu^\prime, \mu_0)d\mu^\prime/(\mu+\mu^\prime) \\
 +\varpi_0H^\text{iso}(\varpi_0, \mu)H^\text{iso}(\varpi_0, \mu_0) -\varpi_0                              -\frac{\varpi_0^2}{2}\big[\mu\log((1+\mu)/\mu)\\
\left. \phantom{P^{(0)}} +\mu_0\log((1+\mu_0)/\mu_0)\big]\right\},    \label{eq-A7}
\end{multline} 
whereas for $m\ge 1$, we substitute the Fourier coefficient that just precedes, viz., 
\begin{equation}
R^{(m)}(\mu, \mu_0)=R^{(m-1)}(\mu, \mu_0), \quad \quad (m\ge 1).   \label{eq-A8}
\end{equation}
We  evaluate  $H^\text{iso}(\varpi_0, \mu)$ and $H^\text{iso}(\varpi_0, \mu_0)$ in Eq.\eqref{eq-A7} using the approximation  formula of Kawabata and Limaye (2011, 2013).%\cite{kaw11, kaw13}. 
 The iteration for successive approximation for $R^{(m)}(\mu, \mu_0)$ is terminated if 
the following condition is satisfied for all combinations of the division points of the quadrature employed for $\mu$ and $\mu_0$:
\begin{equation}
\left|R^{(m)}(\mu, \mu_0)^\text{new}-R^{(m)}(\mu, \mu_0)^\text{old}\right|\le 10^{-7}.   \label{eq-A9}
\end{equation}
The values for the plane albedo $A_\text{pl}(\varpi_0, \mu)$ and the spherical albedo $A_\text{sp}(\varpi_0)$ are then calculated using  those of $R^{(0)}(\mu, \mu_0)$ produced on a square grid of the division points according to 
\begin{subequations}
\begin{align}
A_\text{pl}(\mu, \varpi_0)&=2\int_0^1\!\!R^{(0)}(\mu, \mu^\prime)\mu^\prime d\mu^\prime,  \label{eq-A10a}\\
A_\text{sp}(\varpi_0)\quad &=2\int_0^1\!\!A_\text{pl}(\mu^\prime,\varpi_0)\mu^\prime d\mu^\prime.  \label{eq-A10b}
\end{align}
\end{subequations}
\indent For simplicity,  the Gauss-Legendre quadrature is  employed with $N_\mu=395$
 for Eq.\eqref{eq-A1} and $N_\phi=300$ for Eq.\eqref{eq-A2}. \par
 Our computer code to solve Eq.\eqref{eq-A1} has been tested for the case of conservative isotropic scattering: the maximum relative deviation of the numerical values of reflection function obtained  on a $N_\mu\times N_\mu$ square grid   is  $7.32\times 10^{-7}$ in comparison with those given  by the exact solution:
 \begin{equation}
 R^{(0)}(\mu, \mu_0)=\frac{\varpi_0}{4(\mu+\mu_0)}H^\text{iso}(\mu)H^\text{iso}(\mu_0), \label{eq-A11} 
 \end{equation}
 where the values of  $H^\text{iso}(\mu)$ and $H^\text{iso}(\mu_0)$ are evaluated using  the procedure  discussed in the main text. 
 In addition, the values of the Fourier coefficients  $P^{(0)}(-\mu, \mu_0)$ and $P^{(0)}(\mu, \mu_0)$ of the Henyey-Greenstein phase function (Eq.\eqref{eq-A3}) calculated  by  using Eq.\eqref{eq-A2}
 on this grid   have been checked against those generated  by Eq.\eqref{eq-A4}:
 for $g=0.989$, the maximum relative deviations from the latter results 
  are $3.75\times 10^{-12}$ and $3.23\times 10^{-12}$ for $P^{(0)}(-\mu, \mu_0)$ and $P^{(0)}(\mu, \mu_0)$, respectively, whereas for $g=0.9965$, they are 
   $2.71\times 10^{-11}$ and $3.32\times 10^{-11}$, respectively.\par
 The results for  $A_\text{sp}(\varpi_0)$ obtained using the Henyey-Greenstein phase function with $g=0.99$ and $0.9965$ are shown respectively in columns 2 and 6  of  Table A.1 as functions of $\varpi_0$.  
The $\Delta$-values given in columns 3 and 7 indicate the excess of the last digit figures over those of \cite{rog16}~(RB for short):
\begin{equation}
\Delta\equiv \text{last digit(present)}-\text{last digit(\text{RB})}.   \label{eq-A12}
\end{equation}
Deviations by one unit in the last digit are seen at four locations  in the case of $g=0.9965$
 in contrast to  just one location for $g=0.99$. 
The columns 4 and 8 designated by `Iter' give the  number of iterations required  to solve Eq.\eqref{eq-A1}  for $R^{(0)}(\mu, \mu_0)$ with a given value of $\varpi_0$ respectively for $g=0.99$ and $0.9965$  under the convergence criterion shown  by Eq.\eqref{eq-A9}.  A rapid increase in Iter is clearly seen  when we move   from $\varpi_0=0.9995$ to $0.9999$ especially  in the case of $g=0.9965$, where the number of iterations exceeds $10^4$.  
For reference purpose, the resulting values   for $R^{(0)}(1, 1)$ are given in columns 5 and 9.\par 
  Table A.2 shows  the values of  plane albedo $A_\text{pl}(\mu,$ $ \varpi_0)$ obtained for 6 values of $\varpi_0$ and 14 values of $\mu$ using the  Henyey-Greenstein phase function with $g=0.989$.  The differences $\Delta$ by one unit in the last digit of $A_\text{pl}(\mu)$ are found at 14 locations in comparison with the values  given in Table 2 of \cite{rog16}.
  Also shown  in the bottom row are the corresponding values of the spherical albedo $A_\text{sp}(\varpi_0)$.
They also are in good agreement with those of \cite{rog16} with a one-unit difference in the fourth decimal place found only for  $\varpi_0=0.9995$.\par
In order to make  a further check of the reliability of our procedure  to solve Eq.\eqref{eq-A1} with double peaked phase functions $P(\Theta)$,  we have  also tried the cases with  the two-term Henyey-Greenstein phase function of the form:
\begin{multline}
P(\Theta)=f~P_\text{HG}(g_1; \Theta)+(1-f)~P_\text{HG}(g_2; \Theta),  \\
(g_1\ge 0, ~g_2\le 0, ~f\in [0, 1]).   \label{eq-A13}
\end{multline} 
Following \cite{rog16}, we have employed $g_1=0.995$, $g_2=-0.995$, and $f=0.99$.
With $N_\mu=395$ and $N_\phi=300$, the maximum relative deviations of the values 
of $P^{(0)}(-\mu, \mu_0)$ and $P^{(0)}(\mu, \mu_0)$ obtained by Eq.\eqref{eq-A2} from those given by 
 Eq.\eqref{eq-A4} with $f=0.99$  are found to be $1.87\times 10^{-11}$ and $1.35\times 10^{-11}$, respectively.  \par
The resulting values for $A_\text{pl}(\mu)$, $A_\text{sp}$, and $R^{(0)}(\mu, 1)$ 
obtained for four values of $\varpi_0$, viz., 0.993, 0.997, 0.999, and 0.9995,  are shown  in Table A.3. Also given in the bottom row of Table A.3 are the number of iterations required to get the solution $R^{(0)}(\mu, \mu_0)$ for each value of  $\varpi_0$.  Graphical comparisons of our results for $A_\text{pl}(\mu)$ and $R^{(0)}(\mu, 1)$ with the plots displayed in Fig. 5 and Fig. 6 of \cite{rog16} indicate that the two data sets are in close agreement.\par
 A significant  improvement in execution speed is nevertheless requisite  for our procedure  to be of practical use  for  applications for which  azimuth-angle dependent quantities such as  intensity distributions over  a planetary disk  must be  calculated extensively using highly anisotropic phase functions giving rise to  near-conservative scattering\citep[see, e.g., ][]{yan97}.
%%%%%%%%%%%%%%%%%%%%%%%%%%%%%%%%%%%%%%%%%%%%%%%%%%%%%%%%%%%%%%%%%
%%%%%%%%%%%%%%%%%%%%%%%%%%%%%%%%%%%%%%%%%%%%%%%%%%%%%%%%%%%%%%%%%

%%%%%%%%%%%%%%%%%%%%%%%%%%%%%%%%%%%%%%%%%%%%%%%%%%%%%%%%%%%%%
\clearpage
%%%%%%%%%%%%%%%%%%%%%%%%%%%%%%%%%%%%%%%%%%%%%%
\vspace*{1cm}
{\Large -- Table Captions --}
\begin{enumerate}
\item[Table 1:]~Phase functions employed.$\mbox{}^\dagger$  \\
$\dagger$: ISO: isotropic scattering; LIN: linearly anisotropic
 scattering; \\
RAY: Rayleigh scattering; SOB: \cite{sob75}, \\
Table 7.1; \\
KS3, KS4: \cite{kol72}, Table 1;\\
HUL: \cite{hul80}, Table 29.
\item[Table 2:]~$H(\varpi_0, \mu)$ for conservative isotropic scattering.  
\item[Table 3:]~$H^{(m)}(\varpi_0, \mu)\quad(m=0, 1, 2)$ for conservative Rayleigh scattering. 
\item[Table 4:]~$H^{(m)}(\varpi_0, \mu)\quad(m=0, 1, 2, 3)$ for conservative scattering with the  phase function No.25.
\item[Table A.1:]~The spherical albedos $A_\text{sp}(\varpi_0)$ and the 0-th order Fourier coefficient $R^{(0)}(1,1)$  of reflection function
 calculated for two values of anisotropy parameter $g$   of the Henyey-Greenstein phase function.
\item[Table A.2:]~The  plane albedos $A_\text{pl}(\mu, \varpi_0)$ and spherical albedos $A_\text{sp}(\varpi_0)$ 
calculated  for the Henyey-Greenstein phase function with anisotropy parameter $g=0.989$.
\item[Table A.3:]~The plane albedos $A_\text{pl}(\mu, \varpi_0)$, spherical albedos $A_\text{sp}(\varpi_0)$, and the 0-th order Fourier coefficient 
   $R^{(0)}(\mu, 1)$ of reflection function 
 for the two-term Henyey-Greenstein phase function defined  by Eq.\eqref{eq-A13}
with $g_1=0.995$, $g_2=-0.995$, and $f=0.99$.
\end{enumerate}
\vspace*{1cm}
{\Large -- Figure Captions --}
\begin{enumerate}
\item[Figure 1:]~The phase function No.25.
\item[Figure 2:]~The characteristic functions of the phase function No.25.
\end{enumerate}
%%%%%%%%%%%%%%%%%%%%%%%%%%%%%%%%%%%
 \clearpage
%%%%%%%%%%%%%%%%%%%%%%%%%%%%%%%%%%%%%%%%%%%%%%%%%%%%%%%%%%%%%
\setcounter{table}{0}
\include{kawabata-Table1}

\include{kawabata-Table2}

\include{kawabata-Table3}
\include{kawabata-Table4}
\include{kawabata-TableA1}
\include{kawabata-TableA2}
\include{kawabata-TableA3}
%%%%%%%%%%%%%%%%%%%%%%%%%%%%%%%%%%%%%%%%%%%%%%%
%%%%%%%%%%%%%%%%%%%%%%%%%%%%%%%%%%%%%%%%%%%%%%%
\setcounter{figure}{0}
\onecolumn
%%%%%%%%%%%%%%%%%%%%%%%%%%%%%%%%%%%%%%%%%%%%%%%
\begin{figure}[h]
\hspace*{2cm}\includegraphics[width=14cm]{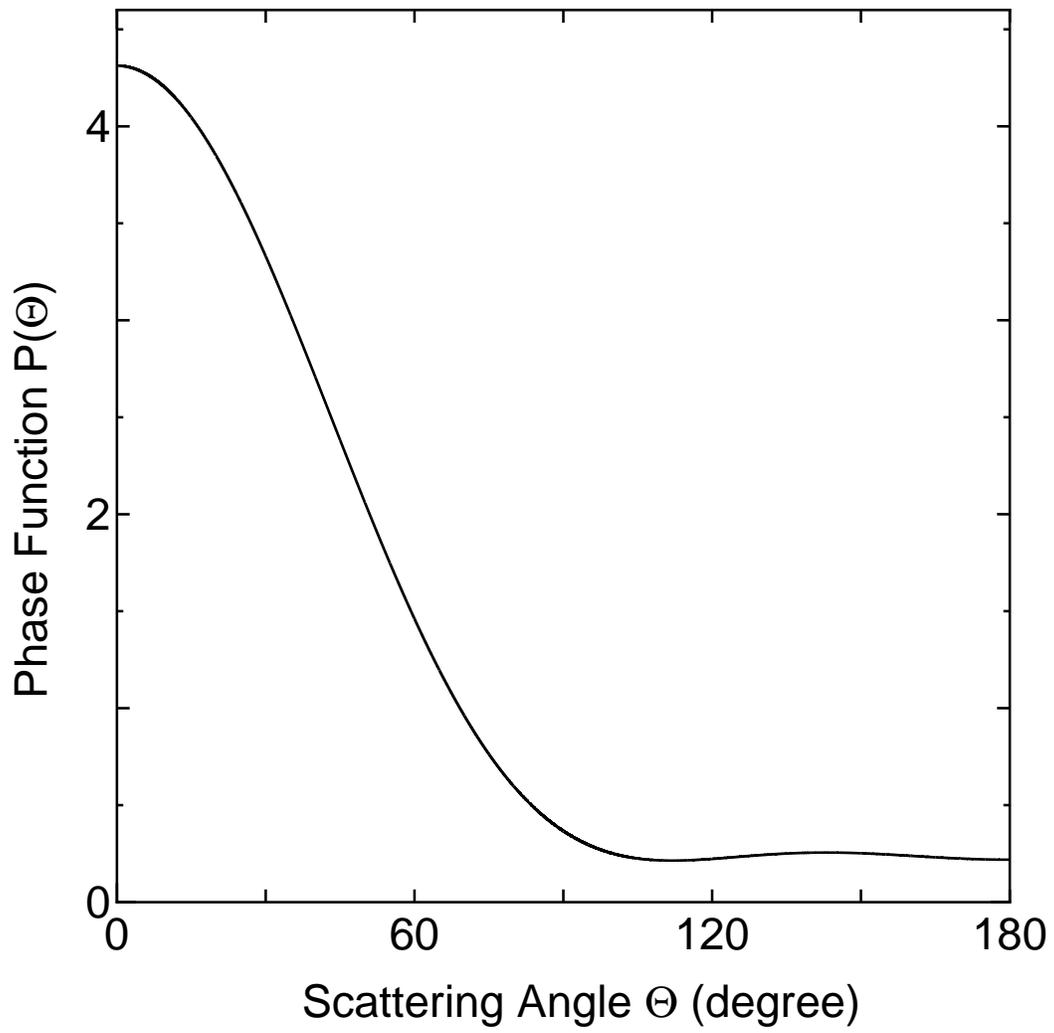}\vspace{8cm}\\
\hspace*{2cm}\caption{The phase function No.25.}
\end{figure}%
\begin{figure}[h]
\hspace*{2cm}\includegraphics[width=14cm]{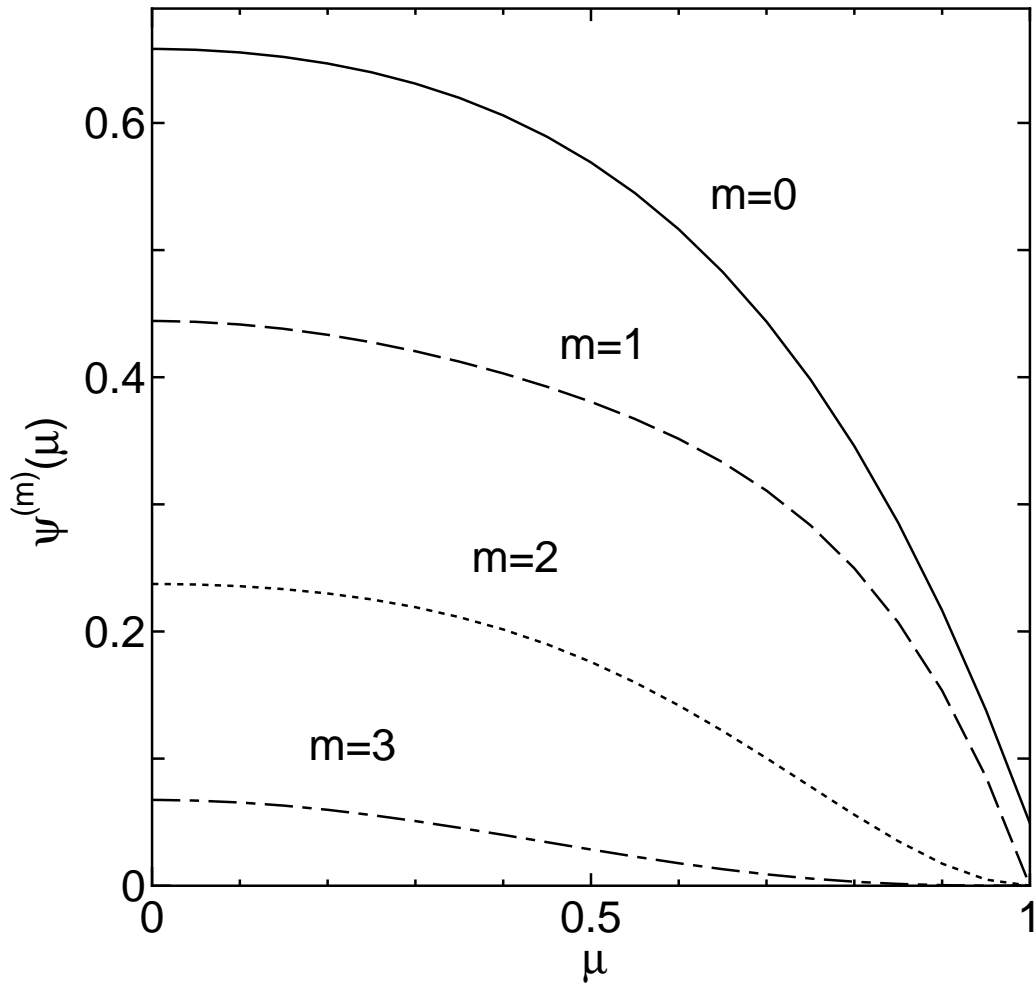}\vspace{8cm}\\
\caption{The characteristic functions of the phase function No.25.}
\end{figure}%
%%%%%%%%%%%%%%%%%%%%%%%%%%%%%%%%%%%%
\clearpage
%%%%%%%%%%%%%%%%%%%%%%%%%%%%%%%%%%%%
\end{document}

%% file: kawabata-Table1.tex
%%%%%%%%%%%%%%%%%%%%%%%%%%%%%%%
\begin{table*}
\begin{center}
%\small
%\hspace*{2cm}\vspace{0.1cm}
\caption{%
The phase functions employed. \label{Tab-1}
}%%
\begin{tabular*}{0.45\linewidth}{@{}crrrrrl@{}}
\tableline\tableline
No. &$M$ &  $x_1$ & $x_2$ & $x_3$ &Old & $\mbox{Ref}^\dagger$ \\
\tableline
1)&0  &  0&  0 &  0 & (1)& ISO  \\
2)&1 &   1  &  0 &    0 &(2)& LIN \\
3)&1  &  0.9 &  0 &   0 & (3)&LIN\\
4)&1  &  0.5&  0&   0 & (4)&LIN\\
5)&1 &  -0.5 &  0 &   0 & (5)&LIN\\
6)&1 &  -0.9 &  0 &   0 & (6)&LIN\\
7)&1 &  -1 &  0 &  0 &(7)&LIN\\
8)&2 &   0 &  0.5 &  0 &(8)& RAY\\   
9)&2 &   1 &  1 &  0& (9)&SOB  \\
10)&2 &   1.5& 1&   0&  (10)&SOB \\
11)&2  &  1.076& 0.795& 0&(11)&KS3\\
12)&2 &   0.240& 0.498 & 0&(12)&KS3\\
13)&2  &  0.092& 0.497& 0&(13)&KS3\\
%14)*&2 &   2.670& 2.470& 0 &(14)&KS3\\
14)&2 &   1.269& 0.909& 0&(15)&KS3\\
15)&2  &  0.566& 0.566& 0&(16)&KS3\\
%17)*&2  &  2.879&2.740&0&(17)&KS3\\
16)&2 &   1.198& 0.869& 0&(18)&KS3\\
17)&2 &   0.540&0.568& 0&(19)&KS3\\
%20)*&2 &   2.560& 2.285 &0&(20)&KS3\\
%21)*&2 &   1.789&1.265& 0&(21)&KS3\\
%22)*&2 &   2.698& 2.459& 0&(22)&KS3\\
%23)*&2 &   1.759&1.283& 0&(23)&KS3\\
18)&3 &  1.006&0.795& 0.215&(24)&KS4\\
19)&3 &  0.208&0.498& 0.098&(25)&KS4\\
20)&3 &  0.083& 0.497& 0.028&(26)&KS4\\
%27)*&3 &  1.972& 2.470& 1.635&(27)&KS4\\
21)&3 &  1.180& 0.909&0.269&(28)&KS4\\
22)&3 &  0.529& 0.566& 0.113&(29)&KS4\\
%30)*&3 &  2.079& 2.740& 1.875&(30)&KS4\\
23)&3 &  1.110&0.869& 0.266&(31)&KS4\\
24)&3 &  0.510&0.568& 0.092&(32)&KS4\\
%33)*&3&1.948& 2.285& 1.432& (33)&KS4\\
25)&3 &  1.615& 1.266& 0.432&(34)&KS4\\
%35)*&3 &  2.028&2.450& 1.569&(35)&KS4\\
26)&3 &  1.560&1.283& 0.494&(36)&KS4\\
27)&3 &  0 & 1&1&(37)&HUL\\
\tableline
 \end{tabular*}
 %% Any table notes must follow the \end{tabular} command.
%\tablenotetext{\dagger}{%
%This table has been reproduced \\
%from Kawabata~(2015).}
\tablenotetext{\dagger}{ISO: isotropic scattering, \\
~LIN: linearly anisotropic scattering, \\
~RAY: Rayleigh scattering, SOB: \cite{sob75}, Table 7.1, \\
~KS3, KS4: \cite{kol72}, Table 1,\\
~HUL: \cite{hul80}, Table 29}.
%\tablenotetext{*}{Phase functions exhibit negative values \\ depending on the values of scattering angle.}
\end{center}
 \end{table*} 
%%%%%%%%%%%%%%%%%%%%%%%%%%%%%%%%%%%%%%%%%%%%%%%%%%%%% 

%% file: kawabata-Table2.tex
  \begin{table*}
 \small
 \caption{$H(\varpi_0, \mu)$ for conservative isotropic scattering.   \ \label{Tab-2}}
 \begin{tabular}{@{}lcccrr@{}}
 \tableline\tableline
 \multicolumn{6}{c}{$m=0$}\\
 \tableline
 $\mu$ & (a) $H(1, \mu)_\text{DE}$ &  (b) $H(1, \mu)_\text{Gauss}$  &(c) $H(1, \mu)_\text{Ana}$   & $\Delta_\text{DE}$ & $\Delta_\text{Gauss}$  \\
 \tableline
 $0     $ & $1.000000000000000$ & $1.000000000000000$ & $1.000000000000000$ & $  0$ & $           0$ \\
 $1(-12)^\dagger$ & $1.000000000014883$ & $1.000000000007192$ & $1.000000000014883$ & $  0$ & $       -7691$ \\
 $1(-11)$ & $1.000000000137316$ & $1.000000000071918$ & $1.000000000137316$ & $  1$ & $      -65398$ \\
 $1(-10)$ & $1.000000001258033$ & $1.000000000719174$ & $1.000000001258033$ & $ 0$ & $     -538859$ \\
 $1(-9) $ & $1.000000011429033$ & $1.000000007191677$ & $1.000000011429033$ & $  0$ & $    -4237356$ \\
 $1(-8) $ & $1.000000102777413$ & $1.000000071910849$ & $1.000000102777413$ & $  0$ & $   -30866564$ \\
 $1(-7) $ & $1.000000912645238$ & $1.000000718519263$ & $1.000000912645238$ & $  0$ & $  -194125975$ \\
 $1(-6) $ & $1.000007975187367$ & $1.000007128565053$ & $1.000007975187366$ & $  1$ & $  -846622314$ \\
 $5(-6) $ & $1.000035852823403$ & $1.000034590885802$ & $1.000035852823402$ & $  1$ & $ -1261937601$ \\
 $1(-5) $ & $1.000068240947974$ & $1.000067211310466$ & $1.000068240947973$ & $  1$ & $ -1029637508$ \\
 $5(-5) $ & $1.000301002209014$ & $1.000300898948308$ & $1.000301002209014$ & $ 0$ & $  -103260706$ \\
 $1(-4) $ & $1.000567416811332$ & $1.000567406348764$ & $1.000567416811332$ & $  0$ & $   -10462568$ \\
 $5(-4) $ & $1.002436861034018$ & $1.002436861035156$ & $1.002436861034018$ & $ 0$ & $        1138$ \\
 $1(-3) $ & $1.004531397798177$ & $1.004531397798826$ & $1.004531397798177$ & $  0$ & $         649$ \\
 $5(-3) $ & $1.018753629227984$ & $1.018753629228115$ & $1.018753629227984$ & $ 0$ & $         131$ \\
 $0.01  $ & $1.034262589374882$ & $1.034262589374948$ & $1.034262589374882$ & $ 0$ & $          66$ \\
 $0.05  $ & $1.136574846838766$ & $1.136574846838780$ & $1.136574846838766$ & $  0$ & $          14$ \\
 $0.10  $ & $1.247350442494436$ & $1.247350442494444$ & $1.247350442494436$ & $ 0$ & $           8$ \\
 $0.15  $ & $1.350833592819941$ & $1.350833592819947$ & $1.350833592819941$ & $ 0$ & $           6$ \\
 $0.20  $ & $1.450351412810095$ & $1.450351412810100$ & $1.450351412810095$ & $  0$ & $           5$ \\
 $0.25  $ & $1.547326233979698$ & $1.547326233979703$ & $1.547326233979698$ & $ 0$ & $           5$ \\
 $0.30  $ & $1.642522264469088$ & $1.642522264469093$ & $1.642522264469087$ & $  1$ & $           5$ \\
 $0.35  $ & $1.736403725419636$ & $1.736403725419642$ & $1.736403725419636$ & $  0$ & $           6$ \\
 $0.40  $ & $1.829275603203368$ & $1.829275603203373$ & $1.829275603203367$ & $  1$ & $           5$ \\
 $0.45  $ & $1.921349591719701$ & $1.921349591719705$ & $1.921349591719701$ & $ 0$ & $           4$ \\
 $0.50  $ & $2.012778769997181$ & $2.012778769997187$ & $2.012778769997181$ & $ 0$ & $           6$ \\
 $0.55  $ & $2.103677409944670$ & $2.103677409944677$ & $2.103677409944670$ & $ 0$ & $           7$ \\
 $0.60  $ & $2.194133019322068$ & $2.194133019322074$ & $2.194133019322067$ & $  1$ & $           6$ \\
 $0.65  $ & $2.284214031328140$ & $2.284214031328147$ & $2.284214031328140$ & $ 0$ & $           7$ \\
 $0.70  $ & $2.373974912536958$ & $2.373974912536965$ & $2.373974912536958$ & $  0$ & $           7$ \\
 $0.75  $ & $2.463459668534999$ & $2.463459668535007$ & $2.463459668534998$ & $ 1$ & $           8$ \\
 $0.80  $ & $2.552704316838003$ & $2.552704316838013$ & $2.552704316838003$ & $ 0$ & $          10$ \\
 $0.85  $ & $2.641738672662855$ & $2.641738672662863$ & $2.641738672662854$ & $ 1$ & $           8$ \\
 $0.90  $ & $2.730587664865337$ & $2.730587664865347$ & $2.730587664865336$ & $ 1$ & $          10$ \\
 $0.95  $ & $2.819272322961027$ & $2.819272322961038$ & $2.819272322961027$ & $ 0$ & $          11$ \\
 $1     $ & $2.907810529078606$ & $2.907810529078616$ & $2.907810529078606$ & $ 0$ & $          10$ \\
 \tableline 
 $\alpha_0$ & $2.000000000000000$ & $2.000000000000003$ & $2.000000000000000$ & $  0$ & $ 3$ \\
 $\alpha_1$ & $1.154700538379251$ & $1.154700538379253$ & $1.154700538379251$ & $  0$ & $ 2$ \\
 $\alpha_2$ & $0.820352482149126$ & $0.820352482149127$ & $0.820352482149125$ & $  1$ & $ 1$ \\
 $\alpha_3$ & $0.637818268031518$ & $0.637818268031519$ & $0.637818268031518$ & $  0$ & $ 1$ \\
 $\alpha_4$ & $0.522227303791946$ & $0.522227303791946$ & $0.522227303791946$ & $  0$ & $ 0$ \\
 \tableline \\
\end{tabular}
\tablenotetext{\dagger}{This is to read as $1\times 10^{-12}.$}
 \end{table*}

%% file: kawabata-Table3.tex
\begin{table*}
 \small
 \caption{$H^{(m)}(\varpi_0, \mu)\quad(m=0, 1, 2)$ for conservative Rayleigh scattering. \ \label{Tab-3}}
 \begin{tabular}{@{}lcccrrrl@{}}
 \tableline\tableline
              &   $m=0$                    &          $m=1$            &            $m= 2$          &      $m=0$        &  $m=1$  &  $m=2$ \\
 \tableline 
 $\mu$ & $H^{(0)}(1, \mu)_\text{DE}$ &   $H^{(1)}(1, \mu)_\text{DE}$  & $H^{(2)}(1, \mu)_\text{DE}$   & $\Delta_\text{Gauss}^{(0)}$ & $\Delta_\text{Gauss}^{(1)}$& $\Delta_\text{Gauss}^{(2)}$  \\
 \tableline
 $0     $ & $1.000000000000000$ & $1.000000000000000$ & $1.000000000000000$ & $           0$ & $  0$ & $           0$ \\
 $1(-12)$ & $1.000000000016596$ & $1.000000000000096$ & $1.000000000002526$ & $        -8652$ & $  0$ & $        -1442$ \\
 $1(-11)$ & $1.000000000153011$ & $1.000000000000960$ & $1.000000000023105$ & $       -73573$ & $  0$ & $       -12262$ \\
 $1(-10)$ & $1.000000001400591$ & $1.000000000009598$ & $1.000000000209464$ & $      -606216$ & $  0$ & $      -101036$ \\
 $1(-9) $ & $1.000000012710707$ & $1.000000000095983$ & $1.000000001878775$ & $     -4767027$ & $  0$ & $      -794504$ \\
 $1(-8) $ & $1.000000114155034$ & $1.000000000959830$ & $1.000000016629077$ & $    -34724889$ & $  0$ & $     -5787476$ \\
 $1(-7) $ & $1.000001012030374$ & $1.000000009598300$ & $1.000000144704046$ & $  - 218391783$ & $  0$ & $    -36398549$ \\
 $1(-6) $ & $1.000008825133414$ & $1.000000095982776$ & $1.000001231173970$ & $   -952451210$ & $  0$ & $   -158740289$ \\
 $5(-6) $ & $1.000039599711670$ & $1.000000479908971$ & $1.000005401461193$ & $  -1419685738$ & $  0$ & $   -236605424$ \\
 $1(-5) $ & $1.000075301735854$ & $1.000000959805674$ & $1.000010153130685$ & $  -1158350605$ & $  0$ & $   -193045569$ \\
 $5(-5) $ & $1.000331283863890$ & $1.000004798538001$ & $1.000043222546018$ & $   -116170238$ & $  0$ & $    -19358094$ \\
 $1(-4) $ & $1.000623662717911$ & $1.000009595851393$ & $1.000079949305436$ & $    -11770005$ & $  0$ & $     -1961907$ \\
 $5(-4) $ & $1.002668234544637$ & $1.000047930490845$ & $1.000324386276762$ & $       1464$ & $  0$ & $         14$ \\
 $1(-3) $ & $1.004951671232060$ & $1.000095739998090$ & $1.000583968656373$ & $        821$ & $  0$ & $         23$ \\
 $5(-3) $ & $1.020374948346988$ & $1.000474001656258$ & $1.002171392904849$ & $        166$ & $  0$ & $          4$ \\
 $0.01  $ & $1.037112197127916$ & $1.000936806673512$ & $1.003706019295775$ & $         85$ & $  0$ & $          2$ \\
 $0.05  $ & $1.146722871265389$ & $1.004303749074392$ & $1.011417375836847$ & $         17$ & $  0$ & $           0$ \\
 $0.10  $ & $1.264709030738373$ & $1.007863449863043$ & $1.017234950896188$ & $          9$ & $  0$ & $           0$ \\
 $0.15  $ & $1.374617126382624$ & $1.010891878239327$ & $1.021334718234983$ & $          6$ & $  0$ & $           0$ \\
 $0.20  $ & $1.480141008914453$ & $1.013514216939769$ & $1.024477614768732$ & $          4$ & $  0$ & $           0$ \\
 $0.25  $ & $1.582856473996785$ & $1.015814563803547$ & $1.026999671573487$ & $          3$ & $  0$ & $           0$ \\
 $0.30  $ & $1.683609020335016$ & $1.017853178392593$ & $1.029084930542283$ & $          3$ & $  0$ & $           0$ \\
 $0.35  $ & $1.782911740143946$ & $1.019675056568940$ & $1.030846544517405$ & $           0$ & $  0$ & $           0$ \\
 $0.40  $ & $1.881101474291584$ & $1.021314804944114$ & $1.032359426590542$ & $           -1$ & $  0$ & $           0$ \\
 $0.45  $ & $1.978411889932651$ & $1.022799649799400$ & $1.033675825564676$ & $           -1$ & $  0$ & $           0$ \\
 $0.50  $ & $2.075011875905088$ & $1.024151403499387$ & $1.034833631367003$ & $           -3$ & $  0$ & $           0$ \\
 $0.55  $ & $2.171027457717663$ & $1.025387805283919$ & $1.035861170412528$ & $           -3$ & $  0$ & $           0$ \\
 $0.60  $ & $2.266555113327543$ & $1.026523466229887$ & $1.036780145601442$ & $           -4$ & $  0$ & $           0$ \\
 $0.65  $ & $2.361670270872751$ & $1.027570553265963$ & $1.037607525713695$ & $           -5$ & $  0$ & $           -1$ \\
 $0.70  $ & $2.456432950827844$ & $1.028539295382113$ & $1.038356806714354$ & $           -7$ & $  0$ & $           0$ \\
 $0.75  $ & $2.550891636732944$ & $1.029438365336758$ & $1.039038880665011$ & $           -8$ & $  0$ & $           0$ \\
 $0.80  $ & $2.645086004697427$ & $1.030275172159541$ & $1.039662650378858$ & $          -10$ & $  0$ & $           0$ \\
 $0.85  $ & $2.739048893857321$ & $1.031056088470063$ & $1.040235474175391$ & $           -8$ & $  0$ & $           0$ \\
 $0.90  $ & $2.832807758015602$ & $1.031786629343648$ & $1.040763494070211$ & $          -12$ & $  0$ & $           0$ \\
 $0.95  $ & $2.926385754186135$ & $1.032471594615600$ & $1.041251882143193$ & $          -14$ & $  0$ & $           0$ \\
 $1     $ & $3.019802571714669$ & $1.033115183226561$ & $1.041705028311013$ & $          -14$ & $  0$ & $           0$ \\
 \tableline 
 $\alpha_0^{(m)}$ & $2.060916240194139$ & $1.021906761813369$ & $1.031716022434768$ & $ -8$ & $ -1$ & $ -2$ \\
 $\alpha_1^{(m)}$ & $1.194021503945255$ & $0.513434581390509$ & $0.518349897878190$ & $ -5$ & $ -2$ & $ -1$ \\
 $\alpha_2^{(m)}$ & $0.849415387249081$ & $0.342956441395375$ & $0.346150939675570$ & $ -3$ & $ -1$ & $ -1$ \\
 $\alpha_3^{(m)}$ & $0.660877538474878$ & $0.257479858818171$ & $0.259828578206846$ & $ -3$ & $ -1$ & $ -1$ \\
 $\alpha_4^{(m)}$ & $0.541342129748284$ & $0.206111902863412$ & $0.207964011044224$ & $ -3$ & $ -1 $ & $ -1$ \\
 \tableline \\
\end{tabular}
 \end{table*}

%% file: kawabata-Table4.tex
 \begin{table*}
 \small
 \caption{$H^{(m)}(\varpi_0, \mu)\quad(m=0, 1, 2, 3)$ for conservative scattering with  the phase function No.25.\ \label{tab-1}}
 \begin{tabular}{@{}lccccr@{}}
 \tableline\tableline
            &   $m=0$                    &          $m=1$            &            $m= 2$          &      $m=3$        &  $m=0$  \\
 \tableline           
 $\mu$ & $H^{(0)}(1, \mu)_\text{DE}$ &   $H^{(1)}(1, \mu)_\text{DE}$  & $H^{(2)}(1, \mu)_\text{DE}$ &  \ \ $H^{(3)}(1, \mu)_\text{DE}$    & $\Delta_\text{Gauss}^{(0)}$ \\
 \tableline
 $0     $ & $1.000000000000000$ & $1.000000000000000$ & $1.000000000000000$ & $1.000000000000000$ & $           0$  \\
 $1(-12)$ & $1.000000000019287$ & $1.000000000012418$ & $1.000000000006485$ & $1.000000000001806$ & $      -10126$  \\
 $1(-11)$ & $1.000000000177711$ & $1.000000000113953$ & $1.000000000059385$ & $1.000000000016505$ & $      -86097$  \\
 $1(-10)$ & $1.000000001625543$ & $1.000000001037235$ & $1.000000000539191$ & $1.000000000149510$ & $     -709408$  \\
 $1(-9) $ & $1.000000014739748$ & $1.000000009349426$ & $1.000000004845330$ & $1.000000001339669$ & $    -5578482$  \\
 $1(-8) $ & $1.000000132240722$ & $1.000000083265026$ & $1.000000042987543$ & $1.000000011842446$ & $   -40635845$  \\
 $1(-7) $ & $1.000001170840152$ & $1.000000730358154$ & $1.000000375217875$ & $1.000000102882019$ & $  -255567028$  \\
 $1(-6) $ & $1.000010192769949$ & $1.000006280675387$ & $1.000003205607351$ & $1.000000873396205$ & $ -1114581633$  \\
 $5(-6) $ & $1.000045667569702$ & $1.000027828706451$ & $1.000014117915755$ & $1.000003823805230$ & $ -1661358993$  \\
 $1(-5) $ & $1.000086774176001$ & $1.000052578726559$ & $1.000026590644233$ & $1.000007179757503$ & $ -1355543584$  \\
 $5(-5) $ & $1.000380953924675$ & $1.000227163368486$ & $1.000113856583786$ & $1.000030467679130$ & $  -135948927$  \\
 $1(-4) $ & $1.000716392706408$ & $1.000423572981826$ & $1.000211270601680$ & $1.000056258291600$ & $   -13772926$  \\
 $5(-4) $ & $1.003055628796602$ & $1.001761518685513$ & $1.000865662249879$ & $1.000227031580472$ & $        2043$  \\
 $1(-3) $ & $1.005661630338316$ & $1.003217294355847$ & $1.001567433056372$ & $1.000407408680368$ & $        1125$  \\
 $5(-3) $ & $1.023195779657177$ & $1.012568948447165$ & $1.005945055909850$ & $1.001498553024185$ & $         229$  \\
 $0.01  $ & $1.042162961133164$ & $1.022156237884193$ & $1.010278472534871$ & $1.002539887685159$ & $         116$  \\
 $0.05  $ & $1.165944061943207$ & $1.077163307543933$ & $1.033205059897492$ & $1.007629711944598$ & $          26$  \\
 $0.10  $ & $1.298996557508615$ & $1.126556721212454$ & $1.051667153593190$ & $1.011335460104873$ & $          14$  \\
 $0.15  $ & $1.422952056128977$ & $1.166117677226941$ & $1.065278863544100$ & $1.013882801955599$ & $          11$  \\
 $0.20  $ & $1.542007295059626$ & $1.199529140748419$ & $1.076059694221327$ & $1.015800442468417$ & $           9$  \\
 $0.25  $ & $1.657940561815204$ & $1.228530008944400$ & $1.084934430647555$ & $1.017317360736430$ & $           6$  \\
 $0.30  $ & $1.771701091285142$ & $1.254142967006965$ & $1.092426420438033$ & $1.018556849542197$ & $           9$  \\
 $0.35  $ & $1.883862487904953$ & $1.277042980842504$ & $1.098866983087174$ & $1.019593577853687$ & $           7$  \\
 $0.40  $ & $1.994799958971425$ & $1.297708580731799$ & $1.104481279622268$ & $1.020476320508766$ & $           7$  \\
 $0.45  $ & $2.104772968606995$ & $1.316495970244862$ & $1.109430070930425$ & $1.021238688226944$ & $           5$  \\
 $0.50  $ & $2.213968530471494$ & $1.333679810910380$ & $1.113832417692899$ & $1.021904791158128$ & $           6$  \\
 $0.55  $ & $2.322525848889461$ & $1.349477613322103$ & $1.117779036517867$ & $1.022492476974057$ & $           6$  \\
 $0.60  $ & $2.430551252736997$ & $1.364065264465195$ & $1.121340639174665$ & $1.023015297612592$ & $           9$  \\
 $0.65  $ & $2.538127703312438$ & $1.377587404825382$ & $1.124573386230724$ & $1.023483762008648$ & $           8$  \\
 $0.70  $ & $2.645321093372461$ & $1.390164638232479$ & $1.127522588778461$ & $1.023906165352140$ & $           6$  \\
 $0.75  $ & $2.752184559704014$ & $1.401898702396663$ & $1.130225299083644$ & $1.024289155814696$ & $           4$  \\
 $0.80  $ & $2.858761518414200$ & $1.412876275723421$ & $1.132712170726780$ & $1.024638132422841$ & $           9$  \\
 $0.85  $ & $2.965087852232256$ & $1.423171842791695$ & $1.135008823692503$ & $1.024957530899517$ & $           8$  \\
 $0.90  $ & $3.071193519202370$ & $1.432849892324790$ & $1.137136865196937$ & $1.025251033165854$ & $          10$  \\
 $0.95  $ & $3.177103757096396$ & $1.441966630766812$ & $1.139114665715679$ & $1.025521723605936$ & $          12$  \\
 $1     $ & $3.282839999426784$ & $1.450571337239516$ & $1.140957957517200$ & $1.025772207444074$ & $          15$  \\
 \tableline 
 $\alpha_0^{(m)}$ & $2.198441980186480$ & $1.305274746410203$ & $1.103528055904102$ & $1.019965669406613$ & $  3$ \\
 $\alpha_1^{(m)}$ & $1.284080546654260$ & $0.684641045400321$ & $0.560755418100964$ & $0.511463360341345$ & $  0$ \\
 $\alpha_2^{(m)}$ & $0.916435068715918$ & $0.464996959699909$ & $0.376039882972427$ & $0.341313769212431$ & $  1$ \\
 $\alpha_3^{(m)}$ & $0.714262503711403$ & $0.352138594610842$ & $0.282860241521951$ & $0.256108303214139$ & $  -1$ \\
 $\alpha_4^{(m)}$ & $0.585708463073474$ & $0.283372856778369$ & $0.226683251412094$ & $0.204943944910463$ & $  0$ \\
 \tableline \\
\end{tabular}
 \end{table*}

%% file: kawabata-TableA1.tex
 \begin{table*}
% \begin{flushleft}
 {\small\bf Table A.1}\quad {\small The spherical albedos $A_\text{sp}(\varpi_0)$ and the 0-th order Fourier coefficient $R^{(0)}(1,1)$  of reflection function  \\
 calculated for two values of anisotropy parameter $g$  of the Henyey-Greenstein phase function.\\}
% {\small\bf Table A.1}\quad {\small Values of  spherical albedo $A_\text{sp}(\varpi_0)$ and 0-th order Fourier coefficient $R^{(0)}(1,1)$ \\ of reflection function
% calculated for two values of anisotropy parameter $g$ and 16 values of $\mu$ \\ of the Henyey-Greenstein phase functions\\}
% {\small\bf Table A.1}\quad {\small Results for Two Cases of $g$ of  the Henyey-Greenstein Phase Function \vspace{0.1cm}} 
\begin{center}
 \small
 \begin{tabular}{@{}llcrrclrrr@{}}
 \tableline\tableline
    & $g=0.99$ &  & & & \phantom{AAAA}& $g=0.9965$& &   &   \\
 \tableline
  \quad$\varpi_0$ & $A_\text{sp}(\varpi_0)$  & $\Delta$   &Iter &$R^{(0)}(1,1)$ &\phantom{AAAAA}  &$A_\text{sp}(\varpi_0)$ & $\Delta$& Iter&$R^{(0)}(1,1)$ \\
 \tableline
 0.9999&  0.795    & 0 &7395 &8.2264(-1) & &0.680     &-1 &10675&6.5501(-1)   \\
 0.9995&  0.604    & 0 &2965 &5.5004(-1) & &0.434     &-1 &5404 & 3.3280(-1) \\
 0.999 &  0.495    & 0 &2295 &4.0830(-1) & &0.318     & 0 &4034 & 2.0442(-1)   \\
 0.997 &  0.310    & 0 &1505 &1.9845(-1) & &0.161     & 0 &2195 & 6.8323(-2)  \\
 0.993 &  0.185    & 0 &941  &8.7064(-2) & &0.815(-1) &-1 &1229 & 2.3824(-2) \\
 0.98  &  0.807(-1)& 0 &459 &2.3829(-2)& &0.308(-1)& -1 &528&6.4558(-3)  \\
 0.97  &  0.558(-1)& 0 &334  &1.4079(-2) & &0.206(-1) & 0 &369  & 4.0191(-3) \\
 0.96  &  0.424(-1)& 0 &263  &9.7493(-3) & &0.154(-1) & 0 & 283 & 2.8936(-3)  \\
 0.95  &  0.340(-1)& 0 &216  &7.3674(-3) & &0.122(-1) & 0 &230  & 2.2482(-3)  \\
 0.94  &  0.283(-1)& 0 &184  &5.8776(-3) & &0.101(-1) & 0 &193  & 1.8306(-3)\\
 0.92  &  0.210(-1)& 0 &141  &4.1314(-3) & &0.743(-2) & 0&146   & 1.3224(-3)    \\
 0.9   &  0.165(-1)& 0 &114  &3.1470(-3) & &0.582(-2) & 0 &116  & 1.0250(-3)   \\
 0.8   &  0.738(-2)& 0 &56   &1.3226(-3) & &0.258(-2) & 0  &57  &4.4663(-4)     \\
 0.7   &  0.431(-1)& 1 &35   & 7.5689(-4)& &0.150(-2) & 0 &38   & 2.5870(-4)   \\
 0.6   &  0.277(-2)& 0 &25   & 4.8187(-4)& &0.966(-3) & 0 &28   &1.6572(-4)    \\
 0.5   &  0.184(-2)& 0 &19   & 3.1937(-4)& &0.644(-3) & 0 &22   &1.1024(-4)    \\
 \tableline
 \end{tabular}
% \newline
% {\footnotesize${}^\dagger$This is to read as $0.815\times 10^{-1}.$} 
% \tablenotetext{}{${}^\dagger$This is to read as $0.807\times 10^{-1}.$} 
 \end{center}
% \end{flushleft}
 \end{table*}

%% file: kawabata-TableA2.tex
 \begin{table*} 
  {\small\bf Table A.2}\quad{\small  The  plane albedos $A_\text{pl}(\mu, \varpi_0)$ and spherical albedos $A_\text{sp}(\varpi_0)$ 
calculated  for the Henyey-Greenstein \\ phase function with anisotropy parameter $g=0.989$.\vspace{0.1cm} }
% {\small\bf Table A.2}\quad{\small  Calculated Values of Plane Albedo $A_\text{pl}(\mu, \varpi_0)$ and Spherical Albedo $A_\text{sp}(\varpi_0)$ \\
% for the Henyey-Greenstein Phase Function with $g=0.989$\vspace{0.1cm} }
 \begin{center}
 \small
 %\hspace*{2cm}\vspace{-3cm}
% \caption{%
% {\bf  Table A.2}\quad Calculated Values of Plane Albedo $A_\text{pl}(\mu, \varpi_0)$ and Spherical Albedo $A_\text{sp}(\varpi_0)$
% }%%
 \begin{tabular}{@{}lrrrrrrrrrrrr@{}}
 \tableline\tableline
 &\multicolumn{11}{c}{$A_\text{pl}(\mu, \varpi_0)$}\\
 \tableline
  \quad$\mu$ & $\varpi_0=$0.99& $\Delta$  & 0.993& $\Delta$  & 0.997& $\Delta$ & 0.999& $\Delta$& 0.9995& $\Delta$  & 0.9999& $\Delta$   \\ \hline
 0.2115(-2)&0.7954& 0& 0.8189& 0& 0.8663& 0& 0.9129& 0& 0.9349& 0& 0.9685& 0 \\
 0.1540(-1)&0.6866& 0& 0.7217& 0& 0.7935& 1& 0.8650& 0& 0.8991& 0& 0.9511& 0 \\
 0.4062(-1)&0.5843& 0& 0.6291& 0& 0.7225& 0& 0.8178& 0& 0.8636& 0& 0.9338& 0 \\
 0.9170(-1)&0.4704& 0& 0.5234& 0& 0.6386& 0& 0.7606& 1& 0.8201& 0& 0.9124& 0 \\
 0.1606     &0.3784& 0& 0.4352& 0& 0.5648& 0& 0.7083& 1& 0.7799& 0& 0.8922& 0 \\
 0.2672     &0.2888& 0& 0.3459& 0& 0.4848& 0& 0.6486& 1& 0.7330& 0& 0.8681& 0 \\
 0.3643     &0.2338& 0& 0.2889& 0& 0.4298& 0& 0.6051& 0& 0.6982& 0& 0.8497& 0 \\
 0.4673     &0.1910& 1& 0.2428& 0& 0.3824& 0& 0.5656& 0& 0.6658& 0& 0.8320& -1 \\
 0.5718     &0.1579& 0& 0.2060& 0& 0.3421& 0& 0.5303& 0& 0.6363& 0& 0.8155& -1 \\
 0.6974     &0.1276& 0& 0.1711& 0& 0.3013& 0& 0.4925& 0& 0.6040& 0& 0.7969& -1 \\
 0.8096     &0.1066& 0& 0.1462& 1& 0.2702& 0& 0.4621& 0& 0.5774& 0& 0.7811& -1 \\
 0.9007     &0.9264(-1)& 0& 0.1292& 0& 0.2479& 0& 0.4393& 1& 0.5571& 0& 0.7687& -1 \\
 0.9645     &0.8423(-1)& 0& 0.1187& 0& 0.2336& 0& 0.4241& 0& 0.5434& 0& 0.7603& -1 \\
 1            &0.7995(-1)& 0& 0.1133& 0& 0.2261& 0& 0.4160& 0& 0.5360& 0& 0.7556& -1 \\
 \tableline
 $A_\text{sp}(\varpi_0)$ & 0.1533& 0& 0.1975& 0& 0.3258& 0& 0.5107& 0& 0.6180& -1& 0.8039& 0 \\
 \tableline
 \end{tabular}
% \tablenotetext{\dagger}{This is to read as $0.2115\times 10^{-2}.$}
 \end{center}
 \end{table*}
 

%% file: kawabata-TableA3.tex
 \begin{table*} 
  {\small\bf Table A.3}\quad{\small The plane albedos $A_\text{pl}(\mu, \varpi_0)$, spherical albedos $A_\text{sp}(\varpi_0)$, and the 0-th order Fourier coefficient \\
   $R^{(0)}(\mu, 1)$ of reflection function 
 for the two-term Henyey-Greenstein phase function defined  by Eq.\eqref{eq-A13}\\
with $g_1=0.995$, $g_2=-0.995$, and $f=0.99$.\vspace{0.1cm} }
% {\small\bf Table A.3}\quad{\small  Calculated Values of Plane Albedo $A_\text{pl}(\mu, \varpi_0)$, Spherical Albedo $A_\text{sp}(\varpi_0)$, and Reflection Function $R^{(0)}(\mu, 1)$ \\
% for a Two-term Henyey-Greenstein Phase Function $f~P_\text{HG}(\Theta;g_1)+(1-f)~P_\text{HG}(\Theta;-g_1)$ with $g_1=0.995$ and $f=0.99$.\vspace{0.1cm} }
 \begin{center}
 \small
 %\hspace*{2cm}\vspace{-3cm}
% \caption{%
% {\bf  Table A.3}\quad Calculated Values of Plane Albedo $A_\text{pl}(\mu, \varpi_0)$, Spherical Albedo $A_\text{sp}(\varpi_0), and Reflection Function $R^{(0)}(\mu, 1)$ for a Three-parameter Phase Function $0.99~P(\Theta,0.995)+0.01~P(\Theta, -0.995)$
% }%%
 \begin{tabular}{@{}lrrrrrrrr@{}}
 \tableline\tableline
% &\multicolumn{11}{c}{$A_\text{pl}(\mu, \varpi_0)$}\\
% \tableline
& $\varpi_0=$0.993& &0.997& & 0.999& &0.9995&      \\ 
\tableline
\quad$\mu$            &$A_\text{pl}(\mu)$ & $R^{(0)}(\mu,1)$ &$A_\text{pl}(\mu)$ & $R^{(0)}(\mu,1)$ & $A_\text{pl}(\mu)$ & $R^{(0)}(\mu,1)$ &$A_\text{pl}(\mu)$ & $R^{(0)}(\mu,1)$ \\
\tableline                
0.2115(-2) & 0.8320&2.4886(-2) &0.8887&3.9946(-2) &0.9370&6.1673(-2) &0.9563&7.4133(-2) \\
0.1540(-1) & 0.7189&4.1566(-2) &0.8120&6.7407(-2) &0.8929&1.0467(-1) &0.9257&1.2603(-1) \\
0.4062(-1) & 0.6273&5.4879(-2) &0.7470&9.0446(-2) &0.8544&1.4183(-1) &0.8986&1.7128(-1) \\
0.9170(-1) & 0.5344&6.7875(-2) &0.6763&1.1504(-1) &0.8101&1.8367(-1) &0.8667&2.2310(-1) \\
0.1606     & 0.4666&7.6763(-2) &0.6192&1.3424(-1) &0.7712&2.1888(-1) &0.8377&2.6782(-1) \\
0.2672     & 0.4074&8.4260(-2) &0.5630&1.5283(-1) &0.7285&2.5569(-1) &0.8044&3.1586(-1) \\
0.3643     & 0.3746&8.9266(-2) &0.5278&1.6539(-1) &0.6986&2.8107(-1) &0.7799&3.4948(-1) \\
0.4673     & 0.3511&9.5343(-2) &0.4999&1.7831(-1) &0.6724&3.0519(-1) &0.7576&3.8096(-1) \\
0.5718     & 0.3343&1.0480(-1) &0.4780&1.9473(-1) &0.6499&3.3158(-1) &0.7376&4.1382(-1) \\
0.6974     & 0.3200&1.2684(-1) &0.4574&2.2767(-1) &0.6268&3.7663(-1) &0.7161&4.6596(-1) \\
0.8096     & 0.3107&1.7385(-1) &0.4429&2.9212(-1) &0.6089&4.5576(-1) &0.6990&5.5190(-1) \\
0.9007     & 0.3048&2.8789(-1) &0.4332&4.3918(-1) &0.5961&6.2602(-1) &0.6862&7.3043(-1) \\
0.9645     & 0.3014&6.5902(-1) &0.4272&8.8606(-1) &0.5878&1.1203~\phantom{(0)} &0.6777&1.2388~\phantom{(0)}     \\
1          & 0.2996&1.5311(~2) &0.4240&1.5494(~2) &0.5833&1.5600(~2) &0.6732&1.5633(~2) \\
 \tableline
 $A_\text{sp}$ & 0.3355& & 0.4738&  & 0.6400&  & 0.7265&   \\
 \tableline
 \text{Iter} & 681 & & 931 & & 1654 & & 2395 &  \\
 \tableline    
 \end{tabular}
% \tablenotetext{\dagger}{This is to read as $0.2115\times 10^{-2}.$}
 \end{center}
 \end{table*}                                                    